\documentclass[12pt]{article}

\usepackage{amssymb, amsmath}
\usepackage{graphicx}
\usepackage{comment}
\usepackage{hyperref}
\usepackage{cite}

\newcommand{\const}{\,{\rm const}\,}

\newtheorem{theorem}{Theorem}
\def\be{\begin{equation}}
\def\ee{\end{equation}}
\def\bea{\begin{eqnarray}}
\def\eea{\end{eqnarray}}

\title{ \bf{Degenerate horizons, Einstein metrics, and Lens space bundles}}

\author{Hari K. Kunduri$^a$\footnote{hkkunduri@mun.ca } \  and James Lucietti$^b$\footnote{j.lucietti@ed.ac.uk } \\ \\
\small \sl $^a$ Department of Mathematics and Statistics, \\  \small \sl Memorial University of Newfoundland,  St John's NL A1C 4P5, Canada
\\ \\ \small \sl $^b$  School of Mathematics and Maxwell Institute of Mathematical Sciences, \\ \small \sl University of Edinburgh,  Edinburgh, EH9 3JZ, UK }

\date{}

\begin{document}

\maketitle


\vskip1.5cm

\begin{abstract}
We present a new infinite class of near-horizon geometries of degenerate horizons, satisfying Einstein's equations for all odd dimensions greater than five. The symmetry and topology of these solutions is compatible with those of black holes. The simplest examples give horizons of spatial topology $S^3\times S^2$ or the non-trivial $S^3$-bundle over $S^2$. More generally, the horizons are Lens space bundles associated to certain principal torus-bundles over Fano K\"ahler-Einstein manifolds.  We also consider the classification problem for Einstein metrics on such Lens space bundles and derive a family which unifies all the known examples (Sasakian and non-Sasakian).
\end{abstract}

\newpage

\tableofcontents
\newpage

\section{Introduction}
The classification of higher dimensional stationary black hole solutions to Einstein's equations is mainly motivated by modern studies of quantum gravity, such as string theory and the gauge/gravity dualities. In this context, extremal black holes are of particular interest due to the fact they do not emit Hawking radiation.

For the sake of being generic, as well as simplicity, in this paper we will consider the vacuum Einstein equations $R_{\mu\nu} =\Lambda g_{\mu\nu}$, where we allow for a cosmological constant $\Lambda$. In four dimensions the black hole uniqueness theorem provides a (partial) answer to the classification problem. However, in higher dimensions, black hole uniqueness is violated. The only explicit black hole solutions known are the spherical horizon topology Myers-Perry black holes~\cite{MP} and the black ring metrics which have $S^1\times S^2$ horizon topology~\cite{ER1, PS}.

Nevertheless, various general results are known, which help constrain the general classification problem. By generalising Hawking's horizon topology theorem~\cite{Hawking} to higher dimensions, Galloway and Schoen~\cite{GS} have shown that the spatial topology of the horizon must be such that it admits a positive scalar curvature metric (i.e. positive Yamabe type). By generalising Hawking's rigidity theorem~\cite{Hawking}, it has also been shown that non-extremal stationary rotating black holes must admit at least $\mathbb{R} \times U(1)$ isometry \cite{rigidity} (for partial results pertaining to extremal rotating black holes see \cite{HIRig}). Both of these topology and symmetry constraints become increasingly weak as one increases the dimensions. Furthermore, there is evidence that black hole uniqueness will be violated much more severely as one increases the dimensions~\cite{blackfolds, DFMRS}. It is clear that in the absence of new ideas, the general classification problem is hopelessly out of reach.

One might expect that more progress could be made by restricting to extremal black holes. To some extent this is the case. The event horizon of all known extremal black holes is a degenerate Killing horizon with compact cross-sections. It turns out that restricting Einstein's equations for a $D$-dimensional spacetime to a degenerate horizon, gives a set of geometric equations for the induced metric $\gamma_{AB}$ on such $(D-2)$- dimensional cross-sections $H$, which depend only on quantities intrinsic to $H$. In the case of vacuum gravity one gets the following equation on $H$
\be
\label{Heq}
R_{AB}= \tfrac{1}{2}h_A h_B -\nabla_{(A} h_{B)} +\Lambda \gamma_{AB}
\ee
where $h_A$ is a 1-form on $H$ (the connection on the normal bundle to $H$), $R_{AB}$ and $\nabla$ are the Ricci tensor and metric connection of $\gamma_{AB}$. By studying solutions to this problem of Riemannian geometry on $H$, one can thus consider the possible horizon geometries independently of the full parent spacetime.  One can understand this feature of degenerate horizons in terms of the near-horizon limit. This limit exists for any spacetime containing a degenerate horizon and allows one to define an associated near-horizon geometry, which must also satisfy the full Einstein equations~\cite{Reall, KLR}.  Classifying near-horizon geometries is then equivalent to classifying solutions to (\ref{Heq}). Because we are ultimately interested in black holes, we will assume that $H$ are compact manifolds.

It turns out that static near-horizon geometries in all dimensions are trivial~\cite{CRT} and hence we will be discussing non-static near-horizon geometries.\footnote{Recall we are discussing only vacuum gravity. If one allows for matter fields, such as Maxwell fields, then non-trivial static near-horizon geometries are possible~\cite{KLstatic}.} Uniqueness of $D=4$ near-horizon geometries has been proved subject to the assumption of axisymmetry~\cite{Haj,LP, KL2}. Indeed this turns out to be the key ingredient for extending the black hole uniqueness theorem to cover the extreme Kerr black hole~\cite{extremeKerr}.  The classification of $D \geq 5$  ($\Lambda=0$) near-horizon geometries has been solved assuming a $U(1)^{D-3}$ rotational isometry \cite{KL1,HI}. The reason for this success ultimately relies on the fact that Einstein's equations for $\Lambda=0$ are integrable for spacetimes with $\mathbb{R}\times U(1)^{D-3}$ isometry\footnote{In fact this structure was not exploited for the $D=5$ analysis of~\cite{KL2}, where the classification problem for $\Lambda \neq 0$ was also considered and reduced to the solution of a single 6th order non-linear ODE.}. For $D=5$ this symmetry assumption is compatible with both asymptotic flatness and Kaluza-Klein asymptotics.
For $D>5$ though this assumption is only compatible with Kaluza-Klein asymptotics.  In the context of asymptotically flat (or globally Anti de Sitter (AdS)) black holes one expects a maximal abelian rotational symmetry group given by the Cartan subgroup of $SO(D-1)$, namely $U(1)^r$ where $r \equiv {\lfloor \tfrac{1}{2} (D-1) \rfloor}$. Hence for $D>5$ we have $r <D-3$, so such cases are not contained in the known classification.

Despite the absence of classification results, a number of near-horizon geometries are known which possess no more than $r$ rotational Killing fields. The Myers-Perry black holes provide a family with spherical horizon topology (including $\Lambda$) \cite{FKLR}. In even dimensions an infinite class has been constructed (including $\Lambda$), which possess cross-sections of the horizon which are $S^2$-bundles associated to $S^1$-bundles over Fano K\"ahler-Einstein manifolds \cite{KL3}. They depend on one continuous angular momentum parameter as well as an integer specifying the precise topology. The simplest examples are cohomogeneity-1 horizon geometries on  $S^2\times S^2$ and $\mathbb{CP}^2 \# \overline{\mathbb{CP}^2 }$.\footnote{These are analogs of the Einstein metrics on complex line bundles over Fano K\"ahler-Einstein manifolds constructed in~\cite{PP} (which generalise Page's Einstein metric~\cite{Page}).}

In this paper we will provide an analogous construction in odd dimensions. In particular we consider horizon cross-section manifolds which are the total space of the associated $S^3$-bundles, and more generally Lens space bundles, to certain principal $T^2$-bundles over Fano K\"ahler-Einstein manifolds. In fact we have already demonstrated an infinite class of near-horizon geometries in this class, which all turn out to possess Sasakain horizon metrics \cite{KL4}. In this sequel we present a more general class of solutions which contain, in  addition to the Sasakian horizons, more generic non-Sasakian geometries. The solutions depend on two angular momenta parameters and certain integers which specify the precise topology. They may be considered as doubly-spinning versions of the Sasakian horizons (which possess only one independent angular momentum)~\cite{KL4} . The simplest examples give horizon geometries on $S^3\times S^2$ and $S^3 \, \tilde{\times} \, S^2$ (the latter, representing the non-trivial bundle, is not allowed in the Sasakian case \cite{KL4}). We emphasise that all the examples we present possess a topology and symmetry compatible with the known constraints for black holes, in particular they are of positive Yamabe type \cite{GS} and are oriented cobordant to $S^{D-2}$ \cite{Reall}.

It is worth remarking that solutions to Einstein's equation have also been of interest in differential geometry~\cite{Besse}. Although space-times correspond to Lorentzian metrics, one can often analytically continue these to complete Riemannian metrics. Indeed, the first example of an inhomogeneous Einstein metric on a compact manifold was found by Page, by taking a certain limit of the Kerr-de Sitter metrics~\cite{Page}, giving a metric on $\mathbb{CP}^2 \# \overline{\mathbb{CP}}^2$. This construction was generalised to five~\cite{Yasui}, and higher dimensions~\cite{CLPP}, by using the Myers-Perry metrics, resulting in an infinite class of inhomogeneous Einstein metrics on $S^3\times S^2$ and $S^3 \, \tilde{\times} \, S^2$, and higher dimensional generalisations.

In fact it turns out that the classification of {\it Einstein} metrics (i.e. solutions to (\ref{Heq}) with $h=0$ and $\Lambda>0$) on $S^3$-bundles and Lens space bundles associated to principal $T^2$-bundles over Fano K\"ahler-Einstein manifolds, is an open problem (of ODE type). A number of examples are known in this class, most notably the Sasaki-Einstein manifolds~\cite{GMSW}, as well as a number of non-Sasaki examples  including those mentioned above \cite{Yasui,LPP, CLPP,DChen}. We show how the classification problem for such metrics can be reduced to a single sixth order non-linear ODE. This allows us to derive all the known examples in a unified form, revealing previously overlooked non-Sasakian examples. These may be of interest in non-supersymmetric generalisations of the AdS/CFT correspondence. 

The organisation of our article is as follows. In section \ref{sec:torusbundle} we provide some mathematical background on the topology and geometry of the class of manifolds we consider, which will also serve to set our notation. In section \ref{sec:Einstein}, we will consider Einstein metrics. In section \ref{sec:horizons} we will give details of the construction of our large class of near-horizon geometries. Sections \ref{sec:Einstein} and \ref{sec:horizons} are presented independently to aid the reader interested in only one of these topics. We also provide an Appendix with some more details.


\section{Principal torus bundles and Lens space bundles}
\label{sec:torusbundle}

In this section we will introduce some mathematical preliminaries  and notation required for our later constructions.

\subsection{Topology}
\label{sec:topology}

Let $K$ be a compact (connected) manifold and $P$ be a  principal $T^2$-bundle over $K$. Such bundles are classified up to isomorphism by the characteristic classes in  $H^2(K, \mathbb{Z}) \oplus H^2(K, \mathbb{Z})$. We can always construct an associated $S^3$-bundle over $K$  as follows. The unit sphere $S^3=\{ |z_1|^2 + |z_2|^2 =1 \; | \; (z_1, z_2) \in \mathbb{C}^2 \}$, so writing an element of $T^2$ as $(e^{i\chi_1}, e^{i\chi_2})$, the natural (left) action of $T^2$  on $S^3$ acts by phase rotation on each $z_i$, that is $(e^{i\chi_1}, e^{i\chi_2})(z_1,z_2)= (e^{i\chi_1}z_1, e^{i\chi_2} z_2)$.  We use this action to define the fibration structure of the $S^3$-bundle in terms of that of the $T^2$-bundle.

More generally we can also construct an associated Lens space bundle over $K$ by using the same (left) $T^2$-action on a Lens space $L(j,k)$. This is possible since $L(j,k)$ is defined by the quotient of $S^3$ by the free (right) action $(z_1, z_2) \sim ( z_1 e^{2\pi i /j}, z_2 e^{2\pi i k/j})$ where $k$ is defined modulo $j$, so $L(j,k) \cong S^3/\mathbb{Z}_j$. We will denote associated sphere and Lens space bundles of this kind by $S= P \times_{T^2} L(j,k)$ (the $S^3$-bundles are given by $j=1$). In this paper we will construct smooth metrics on such manifolds.

We assume $K$ to be a Fano manifold, so $c_1(K)>0$, of complex dimension $n-1$ with a K\"ahler-Einstein structure $( \bar{J}, \bar{g})$. The K\"ahler property means we can represent $c_1(K)$ be the Ricci form $\bar{\rho}$, i.e. $c_1(K)= [ \tfrac{1}{2\pi} \bar{\rho} ]$; then the Einstein property implies the K\"ahler metric $\bar{g}$ must have positive Einstein constant. We will normalise this so that $\text{Ric}(\bar{g})= 2n \bar{g}$. It is well known that $K$ must be simply connected, so $H^2(K,\mathbb{Z})$ has no torsion. This allows one to write its first Chern class $c_1(K) = p a$ where $p$ is the largest positive integer which divides $c_1(K)$, called the Fano index, so $a \in H^2(K, \mathbb{Z})$ is an indivisible class. The Fano index satisfies $p \leq n$ with $p=n$ for $K= \mathbb{CP}^{n-1}$. It is worth noting that in our conventions
\be
\label{Ja}
\bar{J} = 2\pi \left( \frac{p}{2n} \right) a  \; .
\ee

We will work with a subclass of principal $T^2$-bundles over $K$, which are specified by the characteristic classes $(m_1 a, m_2a) \in H^2(K, \mathbb{Z}) \oplus H^2(K, \mathbb{Z})$ where $(m_1, m_2) \in \mathbb{Z}^2$ and we assume at least one of $m_i \neq 0$. We denote such bundles by $P_{m_1, m_2}$ and their associated sphere and Lens space bundles simply by $S$. We will denote the projections of each of these bundles by $\pi_P: P \to K$ and  $\pi_S : S \to K$. To analyse the topology of these manifolds, it is convenient to consider the vector bundle $V= P_{m_1,m_2} \times_{T^2} \mathbb{C}^2$ associated to $P_{m_1,m_2}$, defined by the same left action as above. 

We will now discuss various topological invariants for these bundles. Here we consider the case where the fibre is $S^3$, i.e. $j=1$. Since both the fibre $S^3$ and base $K$ are simply connected, it follows that $S$ also is.\footnote{This can be shown using exactness of the homotopy sequence for $S$.} By standard arguments, as in~\cite{DChen}, one can also show that the second Stiefel-Whitney class of the $S^3$-bundles is given by $w_2(S) = (p+m_1+m_2) \pi_S^* a$ mod$\;2$. It immediately follows that $S$ is a spin manifold if and only if $p+m_1+m_2$ is even.

In the special case $n=2$ we must have $K = \mathbb{CP}^1 \cong S^2$ and so $p=2$. Hence $S$ is the total space of an $S^3$-bundle over $S^2$. Such bundles are classified by $\pi_1(SO(4))= \mathbb{Z}_2$, so there are two inequivalent bundles, one being the trivial one of course. The trivial bundle $S^3 \times S^2$ admits a spin structure, whereas the non-trivial bundle $S^3\, \tilde{\times} \, S^2$ does not. From above we deduce that $S$ is trivial if and only if $m_1+m_2$ is even.

For $n \geq 3$ the sphere bundles $S$ are in fact never trivial, as can be seen using an argument similar to one used in~\cite{GMSW}.  Note $S$ is the unit sphere bundle in $V$, via the natural inclusion. Since $V$ is a complex vector bundle we can form a corresponding real vector bundle, with $SO(4)$ structure group, by ``forgetting" about the complex structure on $V$. The first Pontryagin class of $V$ can then be computed using standard arguments to be $p_1(V)= (m_1^2+m_2^2) a^2$. By assumption $m_i$ are not both vanishing and hence for $n \geq 3$, since $a$ is prop to $\bar{J}$, it must be that $p_1(V)$ is always non-trivial, so the $SO(4)$ bundles $V$ and $S$ are never trivial. Furthermore, for $n \geq 3$, we may define an invariant for such manifolds
\be
\frac{\pi^2 p^2}{n n! V_K} \int_K p_1(V) \wedge \bar{J}^{n-3} =m_1^2+m_2^2  \; ,
\ee
where $V_K$ is the volume of $K$, which shows that two $3$-sphere bundles $S$ can only be homeomorphic if they have the same $m_1^2+m_2^2$.

Later (section \ref{sec:special}) we will consider the topology of Lens space bundles over $K$, i.e. $j>1$, in certain special cases such as when $K$ is a toric manifold.

\subsection{Geometry}
\label{sec:geometry}

Given two-forms $a^i$ for $i=1,2$ on $K$ which represent the classes defining $P$, we may define a principal connection 1-form $\omega^i$ on $P$ with curvature $\Omega^i =d\omega^i=2\pi \pi_P^* a^i$, with respect to a basis $e_i$ for the Lie algebra of $T^2$.\footnote{If $(e_i)$ is a basis of Lie$(T^2)$ then the connection and curvature are the Lie$(T^2)$-valued forms $\omega^i e_i$ and $\Omega^i e_i$.} Denote a basis of vertical vector fields $V_i$ for $i=1,2$, so that $\omega^i(V_j)=\delta^i_j$. Therefore, a principal curvature for $P_{m_1,m_2}$ is given by $\Omega^i= 2\pi \pi_P^*a \, m_i = q^i \pi_P^* \bar{J}$ where
\be
q^i = \frac{2n m_i }{p} \; .
\ee
Now consider a bundle metric on $P_{m_1,m_2}$
\be
\label{torusbundle}
g_P = A^2 \pi_P^*\bar{g} + B_{ij} \omega^i \otimes \omega^j
\ee
where $A$ is a positive constant, $B_{ij}$ is a (constant) positive-definite symmetric $2\times 2$ matrix and $\omega^i$ is the principal connection with curvature $\Omega^i$. Note that $B_{ij}$ induces a bivariant metric on the $T^2$ fibres.\footnote{This makes $\pi_P: (P_{m_1,m_2},g) \to (K,\bar{g})$ a Riemannian submersion with totally geodesic fibers. Such spaces have been well studied~\cite{Besse}.}

A fundamental freedom in the above description is the $GL(2,\mathbb{R})$ automorphism group of the $T^2$ fibres of the principal bundles $P_{m_1, m_2}$.  This acts by $\omega^i \mapsto M^i_{~j} \omega^j$ together with $B \mapsto (M^{-1})^TB (M^{-1})$ (as a matrix), where $M \in GL(2,\mathbb{R})$.  Demanding that $\tfrac{1}{2\pi} \Omega^i$ are integral classes breaks the automorphism group to at least $GL(2,\mathbb{Z})$ (in fact a little more as we will see shortly). This acts on the classes specifying the principal bundle by $m_i \mapsto M^i_{~j} m_j$. Let $m$ be the greatest common divisor of $m_1$ and $m_2$. Bezout's identity states that there always exist integers $r_1, r_2$ such that $m_1r_1+m_2r_2=m$. The $SL(2,\mathbb{Z})$ matrix
\be
\label{M}
M = \left( \begin{array}{cc} r_1 & r_2 \\ -\frac{m_2}{m} &  \frac{m_1}{m} \end{array} \right)
\ee
maps $(m_1, m_2 ) \mapsto (m , 0)$. Hence we may fix the $GL(2,\mathbb{Z})$ automorphism freedom by making such a choice, something we will make use of later. However, this does not completely fix the original $GL(2,\mathbb{R})$ freedom. It is easy see that $(m, 0)$ is preserved by the following group of matrices
\be
\left( \begin{array}{cc} 1 & b \\ 0 & c^{-1} \end{array} \right)
\ee
where $b,c \in \mathbb{R}$ and $c \neq 0$, which is the remaining freedom. This acts on the connection by $(\omega^1, \omega^2) \mapsto (\omega^1+b \omega^2, c^{-1} \omega^2)$. We will not fix the freedom corresponding to $b$ until later. The freedom corresponding to $c$ can be accounted for by rescaling $\omega^2$ without spoiling the curvature since $d\omega^2=0$, which for convenience we will assume henceforth.

We may introduce local coordinates on $P_{m_1,m_2}$ as follows. Locally on $K$ we can always write $\bar{J}= \frac{1}{2} d\bar{\sigma}$ for some 1-form $\bar{\sigma}_a d\bar{x}^a$, where $(\bar{x}^a)$ are coordinates on $K$ with $a=1, \dots 2n-2$. We deduce that $\theta^i= \omega^i - \tfrac{1}{2}q^i\pi_P^*\bar{\sigma}$ are closed 1-forms. Therefore, locally on $P_{m_1,m_2}$ we can write $\theta^i= d\phi^i$, where $\phi^i$ can be identified as coordinates on $T^2=S^1\times S^1$. Hence $(\phi^i, \bar{x}^a)$ are local coordinates on $P_{m_1, m_2}$. In these coordinates
\be
\omega^i = d\phi^i+ \tfrac{1}{2}q^i \bar{\sigma} \qquad \qquad\text{for}\quad  i=1,2
\ee
and the vertical vector fields are the Killing fields
\be
V_i = \frac{\partial}{\partial \phi^i}  \; .
\ee
If we fix the $GL(2,\mathbb{R})$ freedom as above then
\be
\label{q}
q^i=q \delta^i_1 \;, \qquad \qquad q  \equiv \frac{2n m}{p}
\ee
together with $\phi^1 \sim \phi^1+2\pi$ and $\phi^2\sim \phi^2+2\pi c$. The period of $\phi^2$ corresponds to the freedom in rescaling $\omega^2$ mentioned above; also note the remaining gauge freedom $\phi^1 \to \phi^1 + b \phi^2$ is broken to $bc =g \in \mathbb{Z}$ in order to maintain these periodicities.

We can write down a metric on the associated sphere or Lens space bundles $S$ in terms of that on $P_{m_1,m_2}$, as follows. Let $I$ be an open interval in $\mathbb{R}$ and define a metric on $I \times P_{m_1,m_2}$ by $dt^2 +  g_P(t)$, where $t \in I$ and $g_P(t)$ is a 1-parameter family of metrics on $P_{m_1,m_2}$ explicitly given by
\be
g_P(t)= A^2(t)  \bar{g} + B_{ij}(t) \omega^i \otimes \omega^j \;.
\ee
For notational simplicity we have and will suppress the projection map $\pi_P$ henceforth. It turns out to be convenient to make the change of parameterisation of $I$ defined by $t \mapsto x(t)$ where $x'(t) = \sqrt{\det B_{ij}}$. Hence we consider metrics of the form
\be
\label{g}
g = \frac{dx^2}{B(x)} + B_{ij}(x) \omega^i \otimes \omega^j + A^2(x) \bar{g}
\ee
where for notational simplicity we have denoted $B= \det B_{ij}$.
One then extends this to a smooth metric on $S$ by adding the endpoints of $I$ in such a way that $B_{ij}$ has rank-1 at these endpoints.  In other words we ``add" a principal circle bundle at the end points of $I$. The Euler class of these two circle bundles is given by $m_i a$ depending on which $S^1$ factor of $T^2$ collapses. The isometry group of the resulting metrics $g$ is $U(1)^2\times G$, where $G$ is the isometry group of $\bar{g}$ and $U(1)^2$ is generated by the two commuting Killing fields $V_i$. Local coordinates for $S$ are given by $(x,\phi^i, \bar{x}^a)$, where $(\phi^i, \bar{x}^a)$ are the above coordinates on $P_{m_1,m_2}$.

In this paper we will construct smooth metrics on $S$ of the form (\ref{g}).  Examples of Einstein metrics of the above forms on  $S^3$-bundles~\cite{Yasui, DChen} and Lens space bundles~\cite{GMSW}, have been previously constructed.\footnote{Also note that Einstein metrics on $P_{m_1,m_2}$ of the form (\ref{torusbundle}) have been previously constructed~\cite{WZ}.}  We will present the known Einstein metrics~\cite{Yasui, GMSW, DChen} on $S$ in a unified form. We will also construct new degenerate horizons metrics of this form, i.e. solutions to (\ref{Heq}) of the form (\ref{g}).

\subsection{Global analysis}
\label{sec:global}

Here we perform a detailed global analysis of metrics on $I\times P_{m_1,m_2}$ of the form (\ref{g}) and determine the conditions required for them to extend to smooth metrics on Lens space bundles $S$. Due to the form of the explicit local metrics we will derive later, we will work in a $T^2$-basis where (\ref{q}) is satisfied and use the explicit parameterisation
\be
B_{11}= \frac{Q}{q^2 x^{n-1} \Gamma} \, , \qquad \qquad B= \frac{2P}{x^{n-1} \Gamma} \, , \qquad \qquad\Omega= \frac{B_{12}}{B_{11}} \, ,
\ee
so
\be
\label{gansatz}
g= \frac{x^{n-1}\Gamma dx^2}{2P} + \frac{Q}{q^2 x^{n-1}\Gamma}\left(d\phi^1+\frac{q  \bar{\sigma} }{2}+\Omega(x)d\phi^2\right)^2 + \frac{2q^2 P(d\phi^2)^2}{Q} + A^2 \bar{g}
\ee
where $A(x)>0$, $\Gamma(x)>0$, $Q(x)>0$ and $P(x)\geq  0$ for all $x \in [ x_1,x_2 ]$ and $0<x_1<x_2$ are two adjacent simple roots of $P$, i.e. $P(x_1)=P(x_2)=0$, so $I=(x_1,x_2)$. It follows that $P'(x_1)>0$ and $P'(x_2)<0$. We will also assume $\Omega'(x) \neq 0$ for all $x \in [x_1, x_2]$; in particular this implies $\Omega(x_1) \neq \Omega(x_2)$. We will not need any other specific details of the metric and so the result can applied more widely. In particular, later we will apply it to our Einstein metrics in section \ref{sec:Einstein} and horizon metrics in section \ref{sec:horizons}.

For fixed $x_1 < x < x_2$, the metric extends smoothly onto a regular $T^2$-bundle $P_{m_1,m_2} \cong P_{m,0}$ by the construction given above. However,  the Killing fields
\be
\label{app:Ki}
K_i = \frac{1}{\kappa_i} \left( \frac{\partial}{\partial \phi^2} -\Omega(x_i) \frac{\partial}{\partial \phi^1} \right)
\ee
vanish as $x \to x_i$ for each $i=1,2$, where the $\kappa_i$ are normalisation constants which we will choose so that  $(d |K_i| )^2|_{x=x_i}=1$. This gives
\be
\label{app:normKVF}
\kappa_i = q G(x_i)
\ee
where we have defined
\be
\label{app:Gdef}
G(x_i) \equiv  \frac{|P'(x_i)|}{\sqrt{x_i^{n-1} \Gamma(x_i) Q(x_i)}}  \qquad \text{for} \quad i=1,2.
\ee
Demanding that the $K_i$ generate closed orbits with $2\pi$-period is sufficient to ensure smoothness of the fibre with $S^3$ topology.  More generally one can have Lens space fibres, as we show below.

Near each endpoint we define $x- x_i = \frac{2 x_i^{n-1} \Gamma(x_i)}{P'(x_i)} R_i^2$ and find that as $R_i \to 0$, the metric is
\bea
\label{app:gxi}
g &=& [1+O(R_i^2) ] dR_i^2 + [R_i^2+O(R_i^4)] d\xi_i^2 \nonumber \\ &+&\left(  \frac{Q(x_i)}{q^2 x_i^{n-1} \Gamma(x_i)} + O(R_i^2) \right) \left[ d\eta_i + \frac{q \sigma}{2}  +O(R_i^2) d\xi_i \right]^2 + [A(x_i)^2 +O(R_i^2) ]\bar{g}
\eea
where $\xi_i = \kappa_i \phi^2$ and $\eta_i = \phi^1 + \Omega(x_i) \phi^2$. In general one has conical singularities, which are removable if we periodically identify the coordinate $\Delta \xi_i=  2\pi$. Hence we will impose the identification $({\xi}_i, {\eta}_i) \sim ({\xi}_i +2\pi, {\eta}_i )$ in which case the space smoothly approaches $\mathbb{R}^2$ times a principal $U(1)$-bundle over $K$.  However, we must now ensure this identification is compatible with the original $T^2$-bundle our construction is based on, which we now turn to.

Inverting the coordinate change gives for each $i=1,2$
\be
\phi^1= {\eta}_i-\frac{\Omega_i}{\kappa_i} \xi_i \, ,\qquad \qquad \phi^2 = \frac{\xi_i}{\kappa_i}  \; .
\ee
The identification  $({\xi}_i, {\eta}_i) \sim ({\xi}_i +2\pi, {\eta}_i )$ induces the following identification on the original angles:
\be
T_i : (\phi^1, \phi^2) \sim \left( \phi^1- \frac{2\pi \Omega_i}{\kappa_i}, \phi^2+ \frac{2\pi}{\kappa_i} \right)
\ee
for each $i=1,2$ and we have defined $\Omega_i \equiv \Omega(x_i)$ for clarity. These identifications must be compatible with each other and with our original identifications
\begin{equation}
S^{ts} : (\phi^1, \phi^2) \sim (\phi^1+2\pi t, \phi^2+ 2\pi c s)
\end{equation}
where $t,s \in \mathbb{Z}$.  In particular there must be non-zero integers $s_i,t_i$ such that $T_1= S^{t_2 s_2}$ and $T_2 = S^{t_1 s_1}$. These conditions are equivalent to
\be
\frac{\Omega_1}{\kappa_1} = -t_2 \qquad \qquad \frac{\Omega_2}{\kappa_2} = -t_1 \label{app:ti}
\ee
and
\be
\label{c}
c= \frac{1}{\kappa_1 s_2} = \frac{1}{\kappa_2 s_1}  \; .
\ee
If $\text{gcd}(t_2, s_2)>1$ or if $\text{gcd}(t_1, s_1)>1$, this would imply $\xi_1\sim \xi_1 + 2\pi /\text{gcd}(t_2, s_2)$ or $\xi_2 \sim \xi_2 = 2\pi /\text{gcd}(t_1, s_1)$, i.e.  conical singularities at $x=x_1$ or $x=x_2$ respectively. Hence smoothness requires we must have $\text{gcd}(t_1, s_1)= \text{gcd}(t_2, s_2)=1$.   Now, define the integer
\be
\label{app:jdef}
j=s_1t_2 - s_2t_1 \; .
\ee
By combing the above relations one can show $\Omega_2-\Omega_1=c \kappa_1 \kappa_2 j$ and hence since $\Omega_2 \neq \Omega_1$  we deduce that $j \neq 0$ and
\be
\label{sioverj}
\frac{s_i}{j}= \frac{\kappa_i}{\Omega_2-\Omega_1}  \;.
\ee

As mentioned above, the endpoints $x=x_i$ are smooth codimension-2 submanifolds which are each the total space of a $U(1)$-bundle over $K$, as we now show. The metric on these bundles can be read off from (\ref{app:gxi})
\be
g|_{x=x_i}= \frac{Q(x_i)}{q^2 x_i^{n-1}\Gamma(x_i)} \left( d{\eta}_i +\frac{q \bar{\sigma}}{2} \right)^2 + A(x_i)^2 \bar{g} \; .
\ee
The period of $\eta_i$ can be computed as the minimal period consistent with those of $\phi^1$ and $\phi^2$. We find
\be
\Delta \eta_1 = 2\pi \, \text{min}_{n_1, n_2 \in \mathbb{Z}}  \, \left| n_1 + n_2 \Omega_1 c \right| = \frac{2\pi}{s_2} \, \text{min}\, \left|  n_1 s_2  -n_2 t_2 \right| = \frac{2\pi}{s_2}
\ee
where the second equality follows from (\ref{app:ti}) and (\ref{c}) and the third equality from Bezout's identity. Similarly one finds
\be
\Delta\eta_2 = \frac{2\pi}{s_1}  \;.
\ee
We are now in a position to compute the Chern class of each of these $U(1)$-bundles. These bundles have normalised principal connections $A_i =\frac{1}{\Delta \eta_i} (d\eta_i + \frac{ q \bar{\sigma}}{2})$ whose curvatures are
\be
F_1 = \frac{1}{\Delta \eta_1} q \bar{J} = s_2 m a
\ee
and similarly
\be
F_2 = \frac{1}{\Delta \eta_2} q \bar{J} = s_1 m a \; .
\ee
These are integral classes which specify each of the $U(1)$-bundles at the endpoints. Hence we may identify the integers $m_i$ defining the original associated $T^2$-bundle:
\be
\label{app:mi}
m_i =  m s_i \; .
\ee
Notice that $\text{gcd}(m_1,m_2)=m$ implies that we must have $\text{gcd}(s_1, s_2)=1$.

Let us now discuss how the remaining gauge freedom $\phi^1\to \phi^1+ b \phi^2$ acts.  Due to the form of our metric, this can be thought of as shifting $\Omega(x) \to \Omega(x) +b$. However, since $bc=g \in \mathbb{Z}$, this only gives a discrete gauge freedom. In fact, this gauge freedom acts on (\ref{app:ti}) in the following way:
\be
\label{gauge}
t_i \to t_i - g s_i  \; .
\ee
Observe that this leaves $s_i$ and $j$ invariant.

We now show that the fibres of our total space over $K$ are generically Lens spaces. To see this let us write the Killing fields $K_i$ with fixed points at $x=x_i$  (\ref{app:Ki}), in terms of our $2\pi$-periodic normalised basis $(V_1,cV_2)$ of $T^2$ vertical vector fields:
\bea
 \left( \begin{array}{c} K_1 \\  K_2 \end{array} \right) = \left( \begin{array}{cc} t_2 & s_2 \\  t_1 & s_1 \end{array} \right) \left( \begin{array}{c} V_1 \\ cV_2 \end{array} \right)
\eea
where we have used (\ref{app:ti}) and (\ref{c}). The matrix relating the two set of Killing fields is thus
\be
P= \left( \begin{array}{cc} t_2 & s_2 \\  t_1 & s_1 \end{array} \right)
\ee
and notice that $\det P = j$. By a transformation $P \mapsto P A$ where $A \in SL(2,\mathbb{Z})$, which corresponds to a change of $T^2$-basis  $(V_1, c V_2) \mapsto A^{-1} (V_1, cV_2)$, the matrix $P$ can be put in the standard form
\be
\left( \begin{array}{cc} j & k  \\ 0 & 1 \end{array} \right)
\ee
which shows that the fibre metric has Lens space topology $L(j,k)$. This matrix is given by
\be
A= \left( \begin{array}{cc} s_1 &  \hat{t}_1 \\  -t_1 & \hat{s}_1 \end{array} \right)
\ee
where $\hat{s}_1, \hat{t}_1$ are integers such that $s_1 \hat{s}_1+ t_1 \hat{t}_1=1$, which must exist since $\text{gcd}(t_1,s_1)=1$. One can then also read off
\be
\label{k}
k= t_2 \hat{t}_1+ s_2 \hat{s}_1  \; .
\ee
There appears to be a choice of $k$ depending on the integers $(\hat{s}_1, \hat{t}_1)$ one chooses. In fact Euclid's algorithm gives a way of generating all pairs $(\hat{s}_1',\hat{t}_1')$ which satisfy $s_1\hat{s}_1'+t_1\hat{t}_1'=1$, from one pair $(\hat{s}_1,\hat{t}_1)$, by $\hat{t}_1'= \hat{t}_1-n s_1$, $\hat{s}_1'=\hat{s}_1+n t_1$ where $n \in \mathbb{Z}$. This then gives $k'= t_2 \hat{t}_1'+s_2 \hat{s}_1'= k - n j$, so in fact we get a unique $k \; \text{mod} \, j$ and thus a unique Lens space $L(j,k)$.  It is worth noting that under the gauge transformations (\ref{gauge}) the integers $(\hat{s}_1,\hat{t}_1) \mapsto (\hat{s}_1+ g \hat{t}_1, \hat{t}_1)$ and hence $k$ is invariant (as of course must be the case).

To summarise, we have shown that a metric on $I\times P_{m_1,m_2}$ of the form (\ref{gansatz}) extends to a smooth metric on the associated Lens space bundle $S=P_{m_1,m_2} \times_{T^2} L(j,k)$ if there exist non-zero integers $(s_1,s_2, t_1, t_2)$ such that (\ref{app:ti}) and (\ref{sioverj}) are satisfied with $(m_1, m_2, j, k)$ given by (\ref{app:mi}), (\ref{app:jdef}) and (\ref{k}).

Before, moving on it is worth remarking that the compact Riemannian manifolds $(S,g)$ we are considering all admit a contact structure. This is equivalent to the existence of a globally defined 1-form $\eta$ such that $\eta \wedge (d\eta)^n \neq 0$ everywhere on $S$.  For future reference we note that the volume form is
\be
\epsilon = A^{n-1} dx \wedge d\phi^1 \wedge d\phi^2 \wedge \bar{\epsilon}
\ee
where $\bar{\epsilon}$ is the volume form of $K$. Hence $dx \wedge d\phi^1 \wedge d\phi^2$ is a nowhere vanishing $3$-form which defines the orientation on the $L(j,k)$ fibres (this can be checked explicitly by changing to Cartesian coordinates near the endpoints $x=x_i$). Now observe that the Killing field $V_1$ is nowhere vanishing since we are assuming $Q>0$. Its 1-form metric dual
\be
\eta= B_{11}(x) (d\phi^1+ \tfrac{1}{2} q \bar{\sigma} +\Omega(x) d\phi^2)
\ee
is therefore a globally defined nowhere vanishing 1-form on $S$. It is easy to check that
\be
\eta \wedge (d\eta)^n =  - q^{n-1} B_{11}^{n+1} \Omega'  dx \wedge d\phi^1 \wedge d\phi^2  \wedge \bar{J}^{n-1}
\ee
which is everywhere non-vanishing on $S$ by the given assumptions at the start of this section, hence $\eta$ is a contact form.

\subsection{Special cases}
\label{sec:special}
In this section we consider various special cases which allow for a more explicit analysis of the topology of the Lens space bundles we are considering.

\subsubsection{Toric geometry}

In the case $K$ is a toric Fano manifold, we can exploit the contact form $\eta$ on $S$ to determine the topology in more detail, following~\cite{MStoric, MStoric2}.\footnote{We are grateful to James Sparks for explaining to us the constructions used in this section.} Let $C(S)=\mathbb{R}^+\times S$ be the cone over $S$ and $r \in \mathbb{R}^+$ the radial coordinate on the cone. This $2n+2$ dimensional cone has a symplectic form $d(r^2 \eta)$ and moreover is toric, i.e. has a Hamiltonian $T^{n+1}$-action.

For simplicity, consider the five dimensional manifold $S$, so $n=p=2$. In this case the base $K$ must be isometric to the round sphere $S^2$, so
\be
\bar{g}= \frac{1}{4} (d\theta^2 + \sin^2 \theta d\psi^2). \qquad \qquad \bar{\sigma} = \frac{1}{2} \cos \theta d\psi   \; .
\ee
The metric on the total space has a $T^3$-isometry generated by the original $V_1,V_2$ and $\partial / \partial \psi$. It is easy to see that there are exactly four Killing vector fields which vanish on codimension-2 submanifolds and hence they must be linearly related. These are: $K_1, K_2, R_+, R_-$ which vanish at $x=x_1$, $x=x_2$, $\theta=\pi$ and $\theta=0$ respectively, where we have defined
\be
R_{\pm} =  \frac{\partial}{\partial \psi} \pm \frac{m}{2} \frac{\partial}{\partial \phi^1}   \; .
\ee
We now compute the image of the cone under the moment map for the $T^3$-action and verify it is a four faceted polyhedral cone in $\mathbb{R}^3$. Choose a basis $e_i$, with $i=1,2,3$, for $T^3$ and recall the moment map in this basis is $\mu_i = r^2 \eta(e_i)$.   We choose the basis:
\be
e_1 = \frac{\partial}{\partial \phi^1} \qquad e_2 = c \frac{\partial}{\partial \phi^2} \qquad e_3= \frac{\partial}{\partial \psi} +\frac{m}{2} \frac{\partial}{\partial \phi^1}
\ee
where the second term in $e_3$ has been added to ensure the vectors have closed orbits and that the $T^3$-action acts effectively. One then finds that
\bea
\mu|_{x=x_1, \theta= 0}  &=& r^2 B_{11}(x_1) \left( 1, -\frac{t_2}{s_2}, m \right)  \nonumber \\
\mu|_{x=x_1, \theta= \pi} &=& r^2 B_{11}(x_1) \left( 1, -\frac{t_2}{s_2}, 0 \right)    \nonumber \\
\mu|_{x=x_2, \theta= 0} &=& r^2 B_{11}(x_2) \left( 1, -\frac{t_1}{s_1}, m \right)  \nonumber \\
\mu|_{x=x_2, \theta= \pi} &=& r^2 B_{11}(x_2) \left( 1, -\frac{t_1}{s_1}, 0 \right)
\eea
where we have used (\ref{app:ti}) and (\ref{c}). Now, as $r>0$ increases these four vectors in $\mathbb{R}^3$ generate the edges of a four-faceted polyhedral cone. Noting that $B_{11}(x_i)>0$ we deduce that the rays of the cone are generated by the four primitive vectors in $\mathbb{Z}^3$
\be
u_1=[s_2, -t_2, ms_2] \, , \quad u_2= [s_2, -t_2, 0]  \, ,\quad u_3= [ s_1, -t_1, 0] \, , \quad u_4=[ s_1, -t_1, ms_1]  \; .
\ee
For definiteness assume $s_i>0$ and $j>0$. The outward-pointing primitive normals to the four faces are then
\be
v_1 = [-t_2, -s_2, 0] \, , \quad v_2 = [t_1,s_1, 0] \, , \quad v_3=[0,0,-1] \, , \quad v_4= [-m,0,1]  \,  .
\ee
Observe that  the Killing fields corresponding to $v_1,v_2,v_3,v_4$ are (up to signs) the four Killing vectors $K_1, K_2, R_+, R_-$ which vanish of the codimension-2 submanifolds.

It is now readily checked that
\be
m_1 v_1+ m_2 v_2 - j v_3 - j v_4=0 \; .
\ee
This of course coincides with the linear relation between the corresponding four Killing fields $K_i, R_\pm$.
Now assume that $\text{gcd}(m_1,j)= \text{gcd}(m_2,j)=1$. From an analogue of Delzant's theorem given in~\cite{Lerman0}, this implies $C(S) \cong \mathbb{C}^4 // (m_1, m_2, -j, -j )$, where $//$ denotes a symplectic quotient by a $U(1)$ action on $\mathbb{C}^4$ with the specified weights. Hence, restricting to the link $r=1$, we find that
\be
S \cong S^7 // (m_1, m_2, -j, -j )
\ee
where the unit sphere $S^7 \subset \mathbb{C}^4$ and $//$ now denotes the corresponding contact quotient. It now follows by a theorem in~\cite{Lerman} that $\pi_1(S)=0$ and $\pi_2(S)=\mathbb{Z}$. Hence $H_2(S, \mathbb{Z})= \mathbb{Z}$ has no torsion, so by Smale's theorem $S$ must be diffeomorphic to either $S^3\times S^2$ or $S^3 \tilde{\times} S^2$.

We may also use this construction to deduce the second Stiefel-Whitney class $w_2(S)$ as follows. The symplectic cone $C(S) \cong \mathbb{C}^4 // (m_1, m_2, -j, -j )$ possesses a compatible complex structure. The first Chern class is then $c_1(C(S)) = \sum_{a=1}^4 [D_a]$ where $[D_a] \in H^2(S, \mathbb{Z})$ are the Chern classes of the line bundles associated to the four toric divisors $D_a=\{ z_a =0 \}$ where $(z_a)\in \mathbb{C}^4$. In our case $[D_1]= m_1 a$, $[D_2]= m_2 a$, $[D_3]= -j a$ and $[D_4]= -j a$ where $a$ is an indivisible class.  Hence $c_1(C(S))=(m_1+m_2-2j) a$ and therefore  $w_2(S)= (m_1+m_2-2j) a$ mod$\; 2$. In other words $S$ is a spin manifold if and only if $m_1+m_2$ is even for any $j$. From the above we deduce that for all $j$, if $m_1+m_2$ is even $S \cong S^3\times S^2$, whereas if $m_1+m_2$ is odd $S \cong S^3 \tilde{\times} S^2$.

An analogous construction can be given for the higher dimensional $S$ with Fano base $K=\mathbb{CP}^{n-1}$. This again reveals that if $\text{gcd}(m_i,j)=1$ then $\pi_1(S)=0$, $\pi_2(S)=\mathbb{Z}$ and $w_2(S)= (m_1+m_2-n j) a$ mod$\; 2$.

\subsubsection{$U(1)$-bundle total space}
\label{sec:U1bundle}
By re-completing the square we can write the total space metric (\ref{gansatz}) as:
\be
g= B_{22} ( d\phi^2 +  A_B)^2 + g_B
\ee
where
\bea
g_B &=&  \frac{x^{n-1}\Gamma dx^2}{2P}  + \frac{2P}{x^{n-1} \Gamma B_{22}} \left( d\phi^1+ \frac{q \bar{\sigma}}{2} \right)^2  +A^2 \bar{g} \\
A_B &=&  \tilde{\Omega} \left( d\phi^1+ \frac{q \bar{\sigma}}{2} \right)  \label{U1conn}
\eea
and
\be
B_{22}= \frac{2q^2P}{Q}+ \frac{Q \Omega^2 }{q^2x^{n-1} \Gamma}  \qquad \qquad  \tilde{\Omega} \equiv \frac{Q \Omega}{q^2x^{n-1} \Gamma B_{22}}   \; .
\ee
Assume $\Omega(x_i)\neq 0$ for both $i=1,2$, which can always be arranged by shifting $\Omega \to \Omega+b$ appropriately (also recall we have assumed $\Omega_1 \neq \Omega_2$ so they  cannot both vanish). Then $B_{22}(x)>0$ for all $x_1\leq x \leq x_2$ and $\tilde{\Omega}(x_i)=\Omega(x_i)^{-1}$. Hence locally our total space is expressed as $U(1)$ bundle over a base space $B$ with metric $g_B$ and connection $A_B$.

We now consider an interesting special case. Consider the conditions for $B$ to be a smooth manifold. Using $\Delta \phi^1=2\pi$ it turns out that the conical singularities in $g_B$ at $x=x_i$ are removed if and only if  $\Omega_i / \kappa_i = \pm 1$ for both $i=1,2$. In the above language this corresponds to $t_i = \pm 1$; using (\ref{k}) it follows that $k \equiv \pm 1$ mod$\, j$. 
With this choice $B$ is a smooth $S^2$-bundle over $K$ associated to the line bundle over $K$ with Chern class $m a$. This line bundle is $(-\mathcal{L})^{m/p}$ where $-\mathcal{L}$ is the anticanonical bundle over $K$.  The cohomology ring for $B$ in terms of that of $K$ may be written down; in particular $H^2(B,\mathbb{Z}) \cong \mathbb{Z} \oplus H^2(K, \mathbb{Z})$. Since $H^2(K,\mathbb{Z}) \cong \mathbb{Z}^r$ is torsion free, we deduce $H^2(B,\mathbb{Z}) \cong \mathbb{Z}^{r+1}$.

We now turn to the conditions for the total space to be a regular $U(1)$ bundle over $(B,g_B)$. The connection for this bundle is given by  (\ref{U1conn}) and it is easily verified that its curvature $dA_B$ is a smooth two-form on $B$.  A basis of 2-cycles  for $B$ can be given as follows. Let $\Sigma$ be the $S^2$ fibre at a fixed point on the base $K$. Define a section $\sigma: K \to B$ by mapping to one of the poles say $x=x_2$, so $\sigma\Sigma_I$ are 2-cycles in $B$, where $\Sigma_I$ are a basis of 2-cycles in $K$. The 2-cycles in $B$ given by $\Sigma, \sigma \Sigma_I$ then give a natural basis for the free part of $H_2(B, \mathbb{Z})$. We now compute the Chern numbers with respect to this basis, which for general $t_i$ are
\bea
\frac{1}{2\pi c} \int_\Sigma dA_B &=& \frac{1}{c} \left( \frac{1}{\Omega_2}- \frac{1}{\Omega_1} \right) = -\frac{j}{t_1t_2} \\
\frac{1}{2\pi c} \int_{\sigma \Sigma_I} dA_B &=& \frac{m a_I}{\Omega_2c} = -\frac{m_1 a_I}{t_1}
\eea
where $a_I= \int_{\Sigma_I} a$ and the second equalities follow from using (\ref{c}), (\ref{app:ti}), (\ref{app:jdef}) and (\ref{app:mi}). Note that if $t_i = \pm 1$ these are integers, demonstrating regularity of the bundle. In particular $j$ is the Chern number over the $S^2$ fibre of $B$, which gives another way of seeing that the total space is a Lens space bundle $S^3/\mathbb{Z}_j$ over $K$.

We can also show that the total space has a fundamental group $\pi_1(S)= \mathbb{Z}_{\text{gcd}(m_1,j)}$. This can be seen using the homotopy sequence for $S$ viewed as a $U(1)$-bundle over $B$, as follows. Since $\pi_1(B)=0$ we deduce that $\pi_1(S) \cong \mathbb{Z}/ \text{Im} f$ where $f: \pi_2(B) \to \pi_1(S^1)=\mathbb{Z}$. Since $\pi_2(B)= H_2(B,\mathbb{Z})$ then $\text{Im} f = \sum_{i=1}^{r+1} c_i b_i$ where $(b_i) \in \mathbb{Z}^{r+1}$ and $c_i$ are the Chern numbers with respect to a basis of the free part of $H_2(B,\mathbb{Z})$.  In our case we have shown $c_\Sigma=j$ and $c_I= m_1 a_I$ and hence $\text{Im} f = \text{gcd}(m_1,j) \mathbb{Z}$ as claimed.  We deduce that $S$ is simply connected iff $\text{gcd}(m_1,j)=1$, in which case since $m_1t_2- m_2t_1 =mj$ also implies $\text{gcd}(m_2,j)=1$ (note this agrees with the analysis of the toric case in the previous section.)

It is interesting to ask what happens in the general case $t_i \neq \pm 1$ so $B$ is not a smooth manifold. It is easy to see from the above that then one has conical singularities in the fibre of $B$; hence the fibre is an orbifold of $S^2$ with singularities of type $\mathbb{Z}_{t_2}$ and $\mathbb{Z}_{t_1}$ at $x=x_1$ and $x=x_2$ respectively.


\section{Einstein metrics on Lens space bundles}
\label{sec:Einstein}

\subsection{Statement of theorem}
\begin{theorem}
\label{theorem2}
Given any Fano K\"ahler-Einstein manifold $K$, let   $P_{m_1,m_2}$ be the principal $T^2$-bundle over $K$ specified by the characteristic classes $(m_1 a, m_2 a)$ where $a \in H^2(K, \mathbb{Z})$ is the indivisible class given by $c_1(K)=pa$ with $p \in \mathbb{N}$ and $m_i \in \mathbb{Z}$. For all positive integers $(m_1,m_2,j,k)$ such that $m_1>m_2$,  there exists a smooth Einstein metric on the associated Lens space bundles $S= P_{m_1,m_2} \times_{T^2} L(j,k)$.
\end{theorem}
We will prove this in the subsequent sections by an explicit construction of a class of local metrics and then apply the detailed global analysis derived in section \ref{sec:global}. In fact our local metrics are not new. For $K=\mathbb{CP}^{n-1}$, they correspond to the special case of the (Euclideanised) Myers-Perry de Sitter metrics with all rotation parameters equal~\cite{CLPP}. For general $K$, they arise as a special case in the classification of Einstein metrics admitting a closed conformal Killing Yano tensor~\cite{Yasui:2011pr}; they have also been recently found by a different method~\cite{DChen}. Our global analysis encompasses a larger set of solutions than those in~\cite{Yasui, Yasui:2011pr, DChen}, whose analysis restricted attention to $S^3$ fibres ($j=1$). In particular, these works did not capture the infinite class of Sasaki-Einstein manifolds of~\cite{GMSW}, which correspond to the special case $ m_1+m_2= p j$ of the above theorem. 

\subsection{Local construction of metrics}
In this section we consider the classification of  metrics on $I\times P_{m_1, m_2}$ of the form (\ref{g}) which are {\it Einstein}, i.e. satisfy
\be
\text{Ric}(g)= \lambda g.
\ee
The setup and notation is as described in section (\ref{sec:torusbundle}).

To perform the calculations it is convenient to use the following non-coordinate (non-orthonormal) basis of 1-forms
\be
e^0 = \frac{dx}{\sqrt{B(x)}}, \qquad   \qquad e^i = \omega^i , \qquad  \qquad e^a =  \bar{e}^a  \; ,
\ee
where $i=1,2$ and $\bar{e}^a$ is an orthonormal frame for $\bar{g}$. Explicitly the Einstein condition for such metrics is:
\bea
\label{Einstein00}
&&\sqrt{B} (\sqrt{B})'' + \frac{B}{4} B^{lm} B^{pq} B_{lp}' B_{mq}'+ \frac{(2n-2)}{A} \left[ \sqrt{B} (\sqrt{B})' A'+ B A'' \right] +\lambda=0   \\
\label{Einsteinab}
&&(B A')' + \frac{(2n-3)B(A')^2}{A} + \frac{|q|^2}{2A^3} -\frac{2n}{A} +\lambda A =0 \\
\label{Einsteinij}
&&(B  {C^i_{\phantom{i} j}})' + \frac{(2n-2) A' B C^i_{\phantom{i} j}}{A}  -\frac{(n-1)q^iB_{jk}q^k}{A^4} +2\lambda \delta^i_j=0  \; ,
\eea
where $B^{ij}$ denotes the inverse matrix of $B_{ij}$ and we have defined $C^i_{\phantom{i} j} = B^{ik}B_{kj}'$ and $|q|^2 = B_{ij}q^iq^j$. Hence the classification problem reduces to the solution of this system of coupled ODEs for $(A, B_{ij})$. Details of the calculation are given in Appendix \ref{app:curvature}.

It  turns out the above system of ODEs admits a simple first integral. To see this we exploit the $GL(2,\mathbb{Z})$ automorphism group of the $T^2$-fibres of the associated principal bundle $P_{m_1, m_2}$.  As discussed in section (\ref{sec:torusbundle}) we may fix this freedom by setting $(m_1, m_2) \mapsto (m,0)$ where $m= \text{gcd}(m_1, m_2)$, together with $\Delta \phi^1=2\pi$ and $\Delta \phi^2=2\pi c$. It then follows that the $21$ component of (\ref{Einsteinij}) can be integrated to give
\be
C^2_{\phantom{2}1} = \frac{\tilde{r}}{B A^{2n-2}}
\ee
where $\tilde{r}$ is a constant. Since $C^2_{\phantom{2}1} =B_{11}^2 \Omega' /B$ we get
\be
\Omega' =\frac{\tilde{r}}{B_{11}^2 A^{2n-2}}  \; .  \label{1stint}
\ee
where $\Omega \equiv B_{12}/B_{11}$.

The trace of  (\ref{Einsteinij}) simplifies to
\be
\label{Einsteinii}
B''+ \frac{(2n-2)A' B'}{A}  -\frac{(2n-2)|q|^2}{2A^4} +4\lambda =0   \; .
\ee
Now, eliminate $|q|^2$ between equations (\ref{Einsteinab}) and (\ref{Einsteinii}) to find
\be
\label{Pode}
\tilde{P}''= (2n-2)(2n-\lambda A^2) A^{2n-4} - 4\lambda A^{2n-2}   \; ,
\ee
where we have defined
\be
\label{Qdef}
\tilde{P} \equiv B A^{2n-2} \;.
\ee
Thus, given $A(x)$ we can integrate this equation to find $\tilde{P}(x)$ and hence $B(x)$.  In turn this can then be used to deduce $|q|^2$ using either (\ref{Einsteinab}) or (\ref{Einsteinii}), and since $q^i=q \delta^i_1$ this gives $B_{11}$; in particular, equation (\ref{Einsteinab}) gives
\be
\label{gamma11}
B_{11}=\frac{4}{q^2} \left[  nA^2 -\frac{\lambda A^4}{2} - \frac{1}{2 A^{2n-6}} \left( \frac{\tilde{P}A'}{A} \right)' \right]  \; .
\ee
 Then the first integral (\ref{1stint}) can be integrated to give $\Omega$. This is sufficient to give all components of the fibre metric using
 \be\label{fibremet}
\frac{dx^2}{B(x)}+ B_{ij}(x) d\phi^i d\phi^j =  \frac{A^{2n-2} dx^2}{\tilde{P}}  + B_{11} ( d\phi^1 +\Omega d\phi^2)^2 + \frac{\tilde{P}}{ A^{2n-2}B_{11}} (d\phi^2)^2   \; .
\ee
However, in order to satisfy the rest of the Einstein equations $A$ cannot be chosen at will -- we will now derive an ODE which it must satisfy. Equation (\ref{Einstein00}) may be rewritten as
\be
B_{11}^2 \Omega'^2= -B'' - \frac{(2n-2)A'B'}{A}  +  B_{11}' \left( \frac{B}{B_{11}} \right)' - \frac{2(2n-2)A'' B}{A}  -2\lambda  \; .
\ee
We may derive another equation for $B_{11}^2 \Omega'^2$: if one contracts (\ref{Einsteinij}) with $q_iq^j$ one gets
\be
B_{11}^2 \Omega'^2 = \frac{1}{A^{2n-2}} \left( \frac{ A^{2n-2} B B_{11}'}{B_{11}} \right)' -\frac{(n-1)q^2 B_{11}}{A^4} +2\lambda \; .
\ee
Equating these last two expressions and using (\ref{Einsteinii}) gives
\be
\label{gamma11ode}
(A^{2n-2} B_{11}')' + 2(2n-2)A'' A^{2n-3} B_{11}=0  \; .
\ee
Now substituting (\ref{gamma11}) into (\ref{gamma11ode}), and using (\ref{Pode}) to eliminate all multiple derivatives of $\tilde{P}$, finally gives the ODE
\bea
\label{Aode}
&& \tilde{P} (A^2)'''' + \left( 3 \tilde{P}' - \frac{2(n-1)A' \tilde{P}}{A} \right) (A^2)'''  \\  \nonumber &&+ \left(4n(n-2)A(x)^{2(n-2)}+  \frac{(n-2)[(A^2)''\tilde{P}-\tilde{P}'(A^2)']}{A^2} -2(n+3)\lambda A^{2(n-1)} \right) (A^2)''=0  \; .
\eea
Therefore, $(\tilde{P},A)$ satisfy the pair of coupled ODEs (\ref{Pode}) and (\ref{Aode}). Conversely, given a solution to these two ODEs one can construct a solution to the Einstein equations, by defining $B$ and $B_{11}$ from (\ref{Qdef}) and (\ref{gamma11}) and using either equation for $\Omega'^2$ to deduce $\Omega$.\footnote{Then, the first integral (\ref{1stint}) must be automatically satisfied as a consequence of the ODEs (\ref{Pode}) and (\ref{Aode}).}  Therefore we have reduced the classification problem of Einstein metrics of the form (\ref{g}) to solving the pair of non-linear coupled ODEs (\ref{Pode}) and (\ref{Aode}). Notice that in the $n=2$ case, which corresponds to five dimensions, the ODE for $\tilde{P}$ simplifies substantially, and was encountered in the mathematically equivalent classification of 5d near-horizon geometries~\cite{KL1}. It would be very interesting to classify all solutions to this ODE for $\lambda \neq 0$.\footnote{For $n=2$ and $\lambda=0$ one can actually integrate this ODE at least once, so classification is possible depending on the boundary conditions imposed.}

\subsection{A class of explicit Einstein metrics}

\subsubsection{Summary of solutions}
An explicit  family of Einstein metrics is given by
\bea\label{Einsteinmet}
g =  \frac{x^{n-1} dx^2}{2P(x)} + \frac{2q^2P(x)}{Q(x)} (d\phi^2)^2+ \frac{Q(x)}{q^2 x^{n-1}} \left[ d\phi^1+ \frac{q \bar{\sigma}}{2}  + \Omega(x) d\phi^2 \right]^2 +   2x \bar{g}
\eea
where $P(x)$ and $Q(x)$ are the polynomials given by
\be
P(x)=p_0+ p_1 x +  x^{n} -\frac{\lambda x^{n+1}}{n}, \qquad \qquad Q(x)=8(x^n+p_0) \;  ,
\ee
and
\be
\Omega(x)= \omega_0 + \frac{q^2r }{Q(x)}
\ee
where $r$ is a constant defined by
\be
\label{wdef}
r= -\sqrt{\frac{2p_0(p_1n+\lambda p_0)}{n}}.
\ee
The local form of the metrics is thus specified by the parameters $(p_0 , p_1,\omega_0)$ subject to the constraint  $p_0( p_1 n+ \lambda p_0) \geq 0$. The parameter $\omega_0$ is locally redundant and can be set to any value using the remaining $GL(2,\mathbb{R})$ shift freedom $\phi^1 \mapsto \phi^1 +b \phi^2$. However, as described above, we will break the automorphism group of the fibre to $GL(2,\mathbb{Z})$ by choosing $\Delta \phi^1=2\pi$ and $\Delta \phi^2= 2\pi c$; this breaks the remaining gauge freedom to $bc \in \mathbb{Z}$. We will therefore not fix this redundancy until we come to the global analysis of these metrics.

It is worth identifying various notable special cases. The case $p_0=0$ reduces to (choosing $\omega_0=0$)
\be
g =  \frac{dx^2}{2x\left( 1- \frac{\lambda x}{n} + \frac{p_1}{x^{n-1}}\right)} +\frac{q^2}{4} \left( 1- \frac{\lambda x}{n} + \frac{p_1}{x^{n-1}}\right)(d\phi^2)^2 +2x \left[  \frac{4}{q^2} \left(d\phi^1+\frac{q \bar{\sigma}}{2} \right)^2+ \bar{g} \right]
\ee
which, for $K=\mathbb{CP}^{n-1}$ and $q=2$, is nothing but a Schwarzschild metric with cosmological constant and analytically continued time to give Riemannian signature.  If one further sets $p_1=0$ the metric, the singularities at $x=0, n/\lambda$ are removable for $q=2$ and the metric extends onto $0\leq x \leq n/\lambda$, and is locally isometric to the round metric on $S^{2n+1}$.

On the other hand, the case $p_1=0$ is a locally Sasakian-Einstein metric. This can be seen by choosing $\omega_0= \frac{q^2}{4} \sqrt{ \frac{\lambda}{2n}}$ and recompleting the square:
\be
g= \frac{q^2}{4}\left(d\phi^2 + \frac{8x}{q^2}\sqrt{\frac{\lambda}{2n}}( d\phi^1 +\tfrac{1}{2}q\bar{\sigma}) \right)^2  + \left[ \frac{x^{n-1}dx^2} {2P(x)}  + \frac{8P(x)}{q^2x^{n-1}} \left(d\phi^1 +\tfrac{1}{2}q\bar{\sigma} \right)^2 +2x \bar{g} \right]  \; .
\ee
Indeed, it is easy to check that the transverse metric in the square brackets is K\"ahler with K\"ahler form proportional to $d[ x( d\phi^1+ \tfrac{1}{2} q \bar{\sigma})]$.

\subsubsection{Derivation of solutions}

It is clear that $A^2=\alpha x + \beta$ for constants $\alpha,\beta$ is a solution to (\ref{Aode}). By suitable rescaling of $\lambda$ and translation/reflection of $x$, we can fix the constants $\alpha,\beta$ to any values we like: we make the convenient choice $A^2 =2x$. Equation (\ref{Pode}) then  implies $\tilde{P}$ is a polynomial in $x$ given by:
\be
P(x) \equiv \frac{\tilde{P}(x)}{2^n}= \frac{x^{n-1} B}{2} = p_0+ p_1 x +  x^{n} -\frac{\lambda x^{n+1}}{n}
\ee
where $p_i$ are two integration constants. Using (\ref{gamma11}) gives
\be
B_{11} =\frac{4}{q^2} \left[ 2 x (n-\lambda  x) + \frac{2}{ x^{n-1}} (P-x P') \right] = \frac{Q}{q^2 x^{n-1}}
\ee
where we have defined
\be
Q(x)=8( x^n+p_0).
\ee
It now remains to find $\Omega$. The first integral (\ref{1stint}) can be easily further integrated to get
\be
\Omega =  -\frac{q^4 \tilde{r}}{2^{n+2} n Q} +\omega_0
\ee
where $\omega_0$ is an integration constant. Note that the constant $\tilde{r}$ is not free and can be determined from the rest of the equations. One finds
\be
\label{Ksq}
\tilde{r}^2 =  \frac{2^{2n+5}n p_0( p_1 n+ \lambda p_0)}{q^4} \; .
\ee
Notice that $\tilde{r}^2 \geq 0$ implies a constraint on the parameters $p_i$. Without loss of generality we will take $\tilde{r} \geq 0$ (this can always be arranged by redefining $\phi^2 \to -\phi^2$ and $\omega_0 \to -\omega_0$.)  For convenience we write
\begin{equation}\label{Einstein:Omega}
\Omega = \omega_0 + \frac{q^2 r}{Q}
\end{equation}
where we have defined the constant $r$ by (\ref{wdef}).
The remaining equations are identically satisfied.

To summarise, we have completely solved for the local form of the metric (\ref{g}) subject to a single ansatz $A^2 =2 x$. The local form of the metrics is specified by the parameters $(p_0 , p_1,\omega_0)$ subject to the constraint  $p_0( p_1 n+ \lambda p_0) \geq 0$ (this is a consequence of (\ref{Ksq})).

\subsection{Global analysis}

We now turn to the problem of extending the local metrics (\ref{Einsteinmet}) onto smooth, compact Riemannian manifolds, hence we must have $\lambda>0$.   First we note that positive definite signature requires $A^2 \geq 0$ and therefore $x \geq 0$, which in turn implies $P(x) \geq 0$.  Also since $B_{11} \geq 0$  we must have $ x \geq x_* \equiv (-p_0)^{1/n}$. Now consider the potential singularities in the metric as one varies $x$ at a fixed point on $K$. These only occur at $0, x_*, \infty$ and any real roots of $P(x)$: we refer to these collectively as endpoints. A complete manifold requires that $0 \leq x_1<x<x_2$ where $x_1,x_2$ are adjacent endpoints and that the singularities at these endpoints are removable. Compactness requires that the proper distance between the endpoints is finite. Thus, there are a number of cases to consider.

First consider the case where $x_1=0$, i.e. the lower endpoint is at $x=0$. Since $B_{11} = g(V_1, V_1)$, where $V_i = \partial / \partial \phi^i$ are globally defined Killing vector fields, smoothness of the solution implies that we must have $p_0=0$. In this case $B_{22}=g(V_2, V_2)= P(x)/x^n +\omega_0^2 B_{11}$ and hence $p_1=0$ or else $g(V_2,V_2)$ is singular at $x=0$. As mentioned above, the $p_0=p_1=0$ case gives the round metric on $S^{2n+1}$. Henceforth we will assume $p_0 \neq 0$.

Next consider the case where $x_1=x_*>0$ (note that one can never have $x_2 = x_*$). Then we have $p_1=-\lambda p_0/n$ and therefore $P(x)=(1-\frac{\lambda x}{n} ) (x^n-x_*^n)$ and $\Omega=\omega_0$.  Hence the upper endpoint must be given by $x_2= n / \lambda$ and hence $x_*< n/\lambda$.  It follows that $\partial / \partial \phi^1$ and $\partial /\partial \phi^2 -\omega_0 \partial / \partial \phi^1$ vanish at $x=x_*$ and $x=x_2$ respectively; smoothness requires both of these to have closed orbits. Since by construction $\Delta \phi^1=2\pi$ and $\Delta \phi^2= 2\pi c$, we must have $\omega_0 c \in \mathbb{Z}$; hence we may fix the remaining shift freedom $\phi^1\mapsto \phi^1+ b \phi^2$ to set $\omega_0=0$. The metric is now simply
\be
g= \frac{x^{n-1} dx^2}{2(x^n-x_*^n)(1-\frac{\lambda x}{n})} + \frac{8(x^n-x_*^n)}{q^2x^{n-1}} \left(d\phi^1 +\tfrac{1}{ 2}q \bar{\sigma} \right)^2+ \frac{q^2}{4}\left(1-\frac{\lambda x}{n} \right) (d\phi^2)^2+2x \bar{g}  \; .
\ee
There is a conical singularity at $x=x_*$ where $\partial /\partial \phi^1$ vanishes on a codimension-2 submanifold. Smoothness requires this singularity to be absent and noting that $\Delta \phi^1=2\pi$ it is easy to show this corresponds to the condition
\be
q^2 = 4n^2 \left( 1-\frac{\lambda x_*}{n} \right)   \label{qsp}  \; .
\ee
Hence using (\ref{q}) we deduce that $m<p$ and
\be
x_* = \frac{n}{\lambda} \left(1-\frac{m^2}{p^2} \right)  \; .
\ee
At the endpoint $x=x_2$ there is a conical singularity where $\partial / \partial \phi^2$ vanishes on a codimension-2  submanifold. Smoothness at this endpoint is then equivalent to $\Delta \phi^2 =2 \pi c$ where
\be
c^2 = \frac{8x_2}{q^2 \left(1- \frac{x_2^n}{x_*^n} \right)}  \; .
\ee
With these choices the metric $g$ extends to a smooth metric on the compact manifold $S= P_{m,0}\times_{T^2} S^3$ for each $m<p$. This metric was first found in~\cite{LPP}. One could of course take a quotient of the $S^3$-fibre in this case to give a Lens space fibre.

In fact the above is just a special case of the generic case for which $x_1>x_*$, which we now turn to.  In this case we have $x >x_*$ and so $Q>0$. Compactness requires that $x_1 \leq x \leq x_2$ where the endpoints are simple roots of $P(x)$. Also note that in this case the constant $r \neq 0$ and therefore $\Omega' \neq 0$ (if $r=0$ we are back in the case $x_1=x_*$ above). We therefore have a metric exactly of the form studied in section \ref{sec:global}, where conditions were determined in order to extend the metric onto a smooth compact total space $S$ which turns out to be a Lens space bundle over $K$. In particular our Einstein metric is obtained by setting $\Gamma(x)=1$. Using these explicit expressions allows one to write the regularity condition (\ref{sioverj}) in the form
\be
\label{regconds}
\frac{m_i}{j} = X_i(p_0,p_1) \equiv  \left( \frac{p}{2n} \right) \frac{ Q(x_1)Q(x_2) G(x_i)}{8|r| (x_2^n-x_1^n)}  \; ,
\ee
where $m_i$ are the integers specifying the associated $T^2$-bundle $P_{m_1,m_2}$ and $j$ is the integer determining the Lens space fibre $L(j,k)$.

For completeness let us consider the Sasaki special case, which as stated earlier is given by $p_1=0$. Noting that at every root $Q(x_i)= 8 \lambda x_i^{n+1}/n$, allows one to show the quantities $X_i$ defined in (\ref{regconds}) simplify to
\be
X_1 =  \left(\frac{p}{n}\right) \frac{x_2 \left[ n- \frac{(n+1)\lambda x_1}{n} \right]}{x_2-x_1} \; ,  \qquad \qquad
X_2=  -\left(\frac{p}{n}\right) \frac{x_1 \left[ n- \frac{(n+1)\lambda x_2}{n} \right]}{x_2-x_1}
\ee
and therefore $X_1+X_2=p$. Hence
\be
m_1+m_2=jp
\ee
and
\be
\frac{m_1}{m_2} =\frac{x_2\left[ n- \frac{(n+1)\lambda x_1}{n} \right]}{x_1\left[\frac{(n+1)\lambda x_2}{n}-n \right]}  \; .
\ee
It can be shown that in the domain of interest $p/2<X_1<p$ and hence there are an infinite number of solutions $m_1/j$ in the range $p/2<m_1/j<p$.
These regularity conditions are precisely those for the $Y^{j,m_1}$ manifolds~\cite{GMSW, MS}, and also for $K=\mathbb{CP}^{n-1}$ (so $p=n$) reduce to those in~\cite{CLPP} specialised to the cohomogeneity-1 case.

\subsection{Existence of solutions}

We now turn to demonstrating existence of solutions. As shown in the previous section, smoothness of the metric requires the existence of non-zero integers $(m_1,m_2, j)$ which solve $(\ref{regconds})$.
The LHS is a rational number, whereas the RHS is a continuous function of the $(p_0,p_1)$ parameters in the metric. We will therefore need to investigate the functions
\be
X_i(p_0,p_1) =   \left( \frac{p}{2n} \right)  \frac{Q(x_1)Q(x_2) G(x_i)}{8|r|(x_2^n -x_1^n)}  \qquad \text{for} \quad i=1,2.
\ee
Notice that $X_i>0$. It turns out that a more natural parametrisation is in terms of the actual roots $0<x_1<x_2$ of $P(x)$, as we now show.

The existence of the two roots means we can write
\be
P(x)= (x-x_1)(x_2-x) R(x)
\ee
where $R(x)$ is some polynomial of order $n-1$ which we will require to satisfy $R(x)>0$ for all $x \in [x_1, x_2]$.  Writing
\be
R(x)= \sum_{i=0}^{n-1} r_i x^{n-1-i}
\ee
one finds that one can write for all $0 \leq i \leq n-1$
\be
r_i = \frac{\lambda}{n} Y_{i+1} - Y_{i}  \qquad \qquad \text{where} \qquad \qquad Y_i \equiv \frac{x_2^{i}-x_1^{i}}{x_2-x_1}   \; .
\ee
We deduce the original parameters $p_0,p_1$ in terms $x_1,x_2$:
\bea
p_0 &=& -x_1 x_2 \left(  \frac{\lambda}{n}Y_n -Y_{n-1}  \right)  \\
p_1&=& \frac{\lambda}{n} Y_{n+1}-Y_n  \; .
\eea
Therefore we have determined $P(x)$ entirely in terms of the parameters $x_1,x_2$.

We must also determine the bounds on the parameters $(x_1,x_2)$. Clearly they are subject to the restriction $0<x_1<x_2$.  Other constraints which we need to impose are $Q(x_1)=x_1^n+p_0>0$ and $p_0 (p_1 n+ \lambda p_0)> 0$ where $(p_0,p_1)$ are given now explicitly in terms of $(x_1,x_2)$ above. We will see that $R(x)>0$ automatically as a consequence of these and hence there are no more constraints on the parameters.

One can explicitly check
\begin{equation}
Q(x_1) = \frac{(x_2^n - x_1^n)x_1}{(x_2-x_1)}\left[1 - \frac{\lambda x_2}{n}\right]
\end{equation}
so we require $0<x_1<x_2 < n/\lambda$. Furthermore
\be
p_1 n +\lambda p_0= -\frac{n (x_2^n-x_1^n) (1-\frac{\lambda x_1}{n})(1-\frac{\lambda x_2}{n})}{x_2-x_1}  \; .
\ee
Notice that $p_1n+\lambda p_0<0$ automatically and thus we must have $p_0 < 0$. This condition is easily converted to a constraint on the roots using the above formula for $p_0$ resulting in $\frac{\lambda Y_n}{n} > Y_{n-1}$. Notice that this last inequality is simply $r_{n-1}>0$. Using this, we may show that $r_i>0$ for all $i \leq n-1$ and hence deduce that $R(x)>0$, as follows. Simply note that for all $0\leq i <n-1$ we have
\bea
\frac{(x_2-x_1) r_i}{x_1^i} &=& 1-\frac{\lambda x_1}{n} - \left(\frac{x_2}{x_1} \right)^i \left(1-\frac{\lambda x_2}{n} \right) \nonumber \\ &>& 1-\frac{\lambda x_1}{n} - \left(\frac{x_2}{x_1} \right)^n \left(1-\frac{\lambda x_2}{n} \right) = \frac{(x_2-x_1) r_{n-1}}{x_1^{n-1}}
\eea
because $1-\frac{\lambda x_2}{n}>0$ and $x_2/x_1>1$. Hence $r_{n-1}>0$ implies $r_i>0$ for all $i <n-1$ as claimed.

To summarise, the roots $(x_1,x_2)$ must satisfy
\be
\label{xiranges}
0<x_1<x_2< \frac{n}{\lambda} \;,  \qquad \qquad x_2^n -x_1^n > \frac{n}{\lambda} ( x_2^{n-1} -x_1^{n-1})   \;.
\ee
As shown above these conditions are necessary and sufficient for positivity of the metric and for the existence of adjacent simple roots $0<x_1<x_2$ of $P(x)$ such that $P(x)>0$ for $x_1<x<x_2$.
It can be shown that in the domain (\ref{xiranges}) the map $(x_1, x_2) \mapsto (X_1, X_2)$ is onto $\mathbb{R}^+ \times \mathbb{R}^+$. The proof is somewhat tedious and we omit it, although  note it could be deduced from the results of~\cite{DChen} .

Now note the inequalities
\be
\frac{Q(x_2)}{x_2^{n-1}} > \frac{Q(x_1)}{x_1^{n-1}} \, , \qquad \qquad \frac{R(x_1)}{x_1^{n-1}} > \frac{R(x_2)}{x_2^{n-1}}
\ee
where the first follows since $p_0<0$ and the second because $r_i>0$  for all $0\leq i \leq n-1$. Then since
\be
G(x_i)^2 = \frac{(x_2-x_1)^2R(x_i)^2}{x_i^{n-1} Q(x_i)}
\ee
it follows that
\be
G(x_1)>G(x_2)
\ee
and hence
\be
X_1>X_2  \; .
\ee
It follows that for all positive integers $m_1,m_2,j$, such that $m_1>m_2$, there exist a solution to the regularity conditions (\ref{regconds}) such that $(x_1,x_2)$ are functions of the rationals $(m_1/j,m_2/j)$. This establishes theorem \ref{theorem2}.

\section{Construction of near-horizon geometries}
\label{sec:horizons}

\subsection{Statement of main result}
\label{sec:NHequations}
Consider a spacetime containing a smooth degenerate Killing horizon $\mathcal{N}$ of a complete Killing vector field $K$, which admits a cross-section $H$ (i.e. a $(D-2)$-dimensional spacelike submanifold of $\mathcal{N}$ which is intersected once by each integral curve of $K$.) In a neighbourhood of such a horizon we can always introduce Gaussian  null coordinates $(v,r,x^A)$ such that $K = \partial / \partial v$,  $\mathcal{N}=\{ r=0 \}$ and $x^A$ are coordinates on $H$~\cite{IM}. In these coordinates the space-time metric reads
\be
 ds^2 = 2dv \left(dr + rh_A(r,x) dx^A + \tfrac{1}{2} r^2 F(r,x) dv \right)   + \gamma_{AB}(r,x)dx^A dx^B  \; .
\ee
The near-horizon limit \cite{Reall,KLR} is obtained by taking the limit $v \to v/\epsilon, \ r \to \epsilon r$ and $\epsilon \to 0$. The resulting metric is
\be
\label{gNH}
 ds^2 = 2dv \left( dr+ r h_A(x) dx^A + \tfrac{1}{2} r^2 F(x) dv \right) + \gamma_{AB}(x)dx^A dx^B
\ee
where $F, h_A,  \gamma_{AB}$ are a function, a one-form, and a Riemannian metric respectively, defined on the cross-section $H$. We will assume that $H$ is a compact manifold (without boundary).

In this paper we will be interested in finding near-horizon geometry solutions to Einstein's vacuum equations $R_{\mu\nu}=\Lambda g_{\mu\nu}$. We will be mainly focused on $\Lambda \leq 0$, although some of our results will not depend on this.  One can prove (see e.g. ~\cite{KL1}) that these spacetime equations for a near-horizon geometry are in fact equivalent to (\ref{Heq}) and
\be
\label{Feq}
F = \tfrac{1}{2}h_Ah^A - \tfrac{1}{2}\nabla_A h^A +\Lambda  \; ,
\ee
where $R_{AB}$ and $\nabla$ are the Ricci tensor and the covariant derivative of the metric $\gamma_{AB}$, which are equations both defined purely on the cross-section $H$.
For later use it is convenient to note that the contracted Bianchi identity for (\ref{Heq}) is equivalent to
\be
\label{bianchi}
\nabla_A F- F h_A -2h^B\nabla_{[A} h_{B]} -\nabla^B \nabla_{[A} h_{B]} =0 \; .
\ee
We are now ready to state our main result.
\begin{theorem}
\label{theorem1}
Let $P_{m_1,m_2}$ be the principal $T^2$-bundle over any  Fano K\"ahler-Einstein manifold $K$, specified by the characteristic classes $(m_1 a, m_2 a)$ where $m_i \in \mathbb{Z}$ and $a \in H^2(K, \mathbb{Z})$ is the indivisible class given by $c_1(K)=pa$ with $p \in \mathbb{N}$. For a countably infinite set of non-zero integers $(m_1,m_2, j,k)$, there exists a two continuous-parameter family of smooth non-trivial horizon metrics solving (\ref{Heq}) on the associated Lens space bundles $H \cong S= P_{m_1,m_2} \times_{T^2} L(j,k)$.
\end{theorem}
By non-trivial horizon metrics we mean that the 1-form $h$ is not closed, so the corresponding near-horizon geometry is non-static. It is worth noting certain special cases of this theorem. The Sasakian horizon geometries presented in~\cite{KL4} arise as a special case with $m_1+m_2 = pj $. A feature of the Sasakian horizons was that they possess only one independent angular momentum. As we will show, the more general set of horizons presented here, generically possess two independent angular momenta.

\subsection{Near-horizon equations}
Our aim is to construct smooth metrics on $S$ of the form (\ref{g}), i.e.
\be
\label{gamma}
\gamma =\ell^2 \left[ \frac{dx^2}{B(x)} + B_{ij}(x) \omega^i \otimes \omega^j + A^2(x) \bar{g} \right]
\ee
 which solve the horizon equation (\ref{Heq}). Note that for later convenience we have introduced a length scale $\ell$. Since we seek solutions to (\ref{Heq}) we also need to decompose one-forms $h$ on $S$. We will consider 1-forms on $I \times P_{m_1,m_2}$ of the form
\be
\label{h}
h =  -\frac{\Gamma'(x)}{\Gamma(x)} dx + \frac{k_i(x) \omega^i}{\Gamma(x)}
\ee
where $\Gamma(x)$ is some positive function and we use the notation ${}' =\tfrac{d}{dx}$. The 1-form $h$ is invariant under the $U(1)^2\times G$ isometry of $\gamma$, so the associated near-horizon geometry will also have this symmetry.

We now turn to finding solutions $(\gamma, h)$ to the horizon equation (\ref{Heq}) of the form (\ref{gamma}) and (\ref{h}). To perform the calculations it is convenient to use the following non-coordinate (non-orthonormal) basis of 1-forms
\be
e^0 = \frac{\ell dx}{\sqrt{B(x)}}, \qquad   \qquad e^i = \ell \left(d\phi^i+ \frac{q^i \bar{\sigma}}{2} \right) , \qquad  \qquad e^a = \ell \bar{e}^a  \; .
\ee
In this basis we find that the horizon equation (\ref{Heq}) reduces to a set of ODEs. The calculations are routine and are relegated to Appendix \ref{app:curvature}. First note that the $0i$ component of (\ref{Heq}) tells us that
\be
k^i\equiv B^{ij}k_j= \const
\ee
where $B^{ij}$ is the inverse matrix to $B_{ij}$. The rest of the horizon equation (\ref{Heq}) reduces to the following system of ODEs for $(A, B_{ij}, \Gamma)$:
\bea
\label{00eq}
&&\sqrt{B} (\sqrt{B})'' + \frac{B}{4} B^{lm} B^{pq} B_{lp}' B_{mq}'+ \sqrt{B} (\sqrt{B})' \left[ \frac{(2n-2)A'}{A}+\frac{\Gamma'}{\Gamma} \right] \nonumber  \\ && \qquad + B \left[ \frac{\Gamma''}{\Gamma} -\frac{\Gamma'^2}{2\Gamma^2}+ \frac{(2n-2) A''}{A} \right] +\lambda=0   \\
\label{abeq}
&&(B A')' +B A' \left[  \frac{\Gamma'}{\Gamma} + \frac{(2n-3)A'}{A} \right] + \frac{|q|^2}{2A^3} -\frac{2n}{A} +\lambda A =0 \\
\label{ijeq}
&&(B  {C^i_{\phantom{i} j}})' + B C^i_{\phantom{i} j}\left[ \frac{(2n-2) A'}{A} + \frac{\Gamma'}{\Gamma} \right] +\frac{k^ik_j}{\Gamma^2} -\frac{2(2n-2)q^iB_{jk}q^k}{4A^4} +2\lambda \delta^i_j=0  \; ,
\eea
which correspond to the $00$, $ab$ and $ij$ components of (\ref{Heq}) respectively, where for convenience we have defined $C^i_{\phantom{i} j} = B^{ik}B_{kj}'$, $|q|^2=B_{ij} q^iq^j$ and $\lambda \equiv \Lambda \ell^2$. Note that the trace of (\ref{ijeq}) gives:
\be
\label{traceeq}
B'' + B' \left[ \frac{(2n-2)A'}{A} +\frac{\Gamma'}{\Gamma} \right] + \frac{|k|^2}{\Gamma^2} -\frac{(2n-2)|q|^2}{2A^4} +4\lambda =0
\ee
where $|k|^2 =B_{ij} k^i k^j$.

We may write down a first integral of the above system. The $0$ component of (\ref{bianchi}) can be integrated to give
\be
\label{firstint}
F= \frac{a_0\Gamma + |k|^2}{\ell^2\Gamma^2}
\ee
where $a_0$ is the integration constant.  Equation (\ref{Feq}) can now be simplified using (\ref{firstint}) to give the promised first integral\footnote{In principle one could check that $a_0$ is indeed a constant directly from our system of ODEs (\ref{00eq}), (\ref{abeq}), (\ref{ijeq}), although this would be a rather cumbersome calculation.}
\be
\label{a0}
a_0 = -\frac{|k|^2}{2\Gamma}+\frac{1}{2}\nabla^2 \Gamma+\lambda \Gamma \; ,
\ee
where for any function depending only on $x$ we have
\be
\nabla^2 f \equiv  (B f')' +\frac{2(n-1) B A'f'}{A}  \; .
\ee
We are interested in non-static near-horizon geometries, so $|k|>0$, with compact horizon cross-sections $H$. For definiteness  we will focus on $\lambda \leq 0$, although many of our results are valid for  (sufficiently small) $\lambda>0$. Therefore, integrating (\ref{a0}) over $H$ shows that $a_0<0$, which we will assume henceforth.

The significance of the constant $a_0$ is revealed by changing coordinate $r \to \Gamma(x) r$ in the full near-horizon geometry resulting in
\be
ds^2 = \Gamma(x) \left[ \frac{a_0 r^2 dv^2}{\ell^2}  +2dvdr \right] + \ell^2 \left[ A(x)^2 \bar{g}+ \frac{dx^2}{B(x)} + B_{ij}(x) \left( \omega^i + \frac{k^i rdv}{\ell^2} \right)\left(\omega^j +\frac{k^j rdv}{\ell^2} \right)\right]
\ee
which demonstrates that all near-horizon geometries in this class are fibrations over AdS$_2$, which inherit the AdS$_2$ isometry group $SO(2,1)$.

For later reference we note that there are certain redundancies in our parametrisations. One is a trivial scaling freedom associated to the constant $\ell$ we introduced:
\be
\label{scaling}
(x, A, B_{ij}, \ell) \mapsto (s^6 x, s A, s^2 B_{ij}, s^{-1} \ell)
\ee
with $\Gamma, k_i$ fixed, where $s \neq 0$ is any constant. Another is a trivial scaling freedom in the definition of $(\Gamma, k_i)$ which acts as
\be
(\Gamma, k_i, a_0) \mapsto (t\Gamma, tk_i,  t a_0)
\ee
with all else fixed, where $t$ is a positive constant.

It appears that solving the above system of ODEs is a very difficult problem. In fact this is also unsolved in the Einstein case ($h=0$), which we considered in section \ref{sec:Einstein}. Nevertheless, we can still find explicit solutions of interest, as we show in the next section.

\subsection{A class of explicit near-horizon geometries}

We wish to find non-trivial solutions to our system of ODEs. The known examples with $h\neq 0$ correspond to the Myers-Perry solutions (see Appendix \ref{app:MP}) and the Sasakian horizons~\cite{KL4}. If one sets $h \equiv 0$, then the problem reduces to finding Einstein metrics in the same class, which we considered in section \ref{sec:Einstein}. For reference, in our notation the Einstein case corresponds to $\Gamma=1$, $k^i=0$ and therefore from equation (\ref{a0}) also $a_0=\lambda$.

We will consider the following  ansatz $A(x)^2 = \alpha^2 x$
for constant $\alpha>0$, which contains all the aforementioned examples. At this stage we exploit the scaling freedom (\ref{scaling}) to fix $\alpha^2$ to a convenient value -- we will make the choice $\alpha^2=2$ (achieved by $s^2=2\alpha^{-1}$),
so
\be
\label{A}
A(x)^2=2x
\ee
which will allow us to connect to the Sasakian case most easily~\cite{KL4}. We will supplement this by an ansatz for $\Gamma$
\be
\label{Gamma}
\Gamma(x) = \beta x + \xi
\ee
where $\beta, \xi$ are constants, which again contains all the known cases. Hence it remains to determine the matrix $B_{ij}(x)$.
We will present solutions within this ansatz which generalise the known cases.

\subsubsection{Summary of solutions}

In this section we present our solutions, leaving their derivation to the next section in order to not obscure our results.

We have found a set of solutions to (\ref{Heq}) given by
\begin{eqnarray}
\gamma &=&\ell^2 \left[ \frac{x^{n-1}\Gamma dx^2}{2P} + \frac{Q}{q^2 x^{n-1}\Gamma}\left(d\phi^1+\frac{q  \bar{\sigma} }{2}+\Omega(x)d\phi^2\right)^2 + \frac{2q^2 P(d\phi^2)^2}{Q} + 2x \bar{g} \right] \label{gammasol} \\
 h &=&\frac{1}{\ell^2} \left[  \frac{k^i}{\Gamma} \frac{\partial}{\partial \phi^i} - \frac{2P \Gamma'}{x^{n-1} \Gamma^2} \frac{\partial}{\partial x} \right]   \label{hsol}
\end{eqnarray}
where for simplicity we have expressed the 1-form $h_A$ as a vector field $h^A=\gamma^{AB}h_B$. The functions and constants appearing in the above metric are defined as follows: $\Gamma(x)$ is given by (\ref{Gamma}),
\bea
\label{P}
P(x) &=& -\frac{\beta \lambda x^{n+2} }{(n+1)}  + \frac{\left[ n(n-1) \beta+ a_0 -\lambda \xi (n+2) \right]x^{n+1}}{n(n+1)}   +\xi x^n +p_1  x +p_0\\
\label{Q}
Q(x) &=&  \frac{8}{n+1} \left[  v_4 x^{n+1}  + \xi(n+1) x^n + (n+1)p_0\right]  \\
\Omega(x) &=& \omega_0 +  \frac{q^2 R(x)}{Q(x)} \label{Omega} \\
R(x) &=& \sqrt{\frac{2w}{n}} \frac{(1+n)\beta p_0 - v_1\xi x^n}{v_4}  \; , \label{Rsum}
\eea
and
\bea
v_0 &=&  n\beta -a_0  +\lambda \xi  \label{v0}  \\
v_1 &=& (n-1)\beta +  \lambda \xi - a_0 \label{v1} \\
v_2 &=& (n+1)v_0 +\lambda \xi \label{v2} \\
v_3 &=& 2 n \beta + 2 \lambda \xi - a_0 \label{v3} \\
v_4 &=& 2n\beta +\lambda \xi -a_0 \label{v4} \\
p_1 &=& \frac{\beta v_2 p_0}{n\xi v_1} \label{p1} \\
w &=& \frac{v_0v_3}{\xi v_1} \label{w}  \; .
\eea
The quantities $k^1,k^2$ are constants given by
\bea
k^2 &=&  \frac{2\epsilon \sqrt{2 \xi v_0}}{q} \label{k2} \\
k^1 &=& -\omega_0 k^2 - \frac{\epsilon\,  \text{sgn}(v_0) q\beta v_1}{2v_4} \sqrt{ \frac{v_3 n}{v_1} }\label{k1}
\eea
where $\epsilon=\pm 1$. The solution is parametrised by the constants $(\ell, a_0, \beta, \xi, p_0, \omega_0)$, subject to the constraints
\bea
&&\xi v_0 \geq 0 \label{constr1} \\
&&\xi v_1 \neq 0 \label{constr2}\\
&&w \geq 0 \label{constr3}\\
&&v_4 \neq 0  \label{constr4} \; .
\eea
Recall $\lambda= \Lambda \ell^2$ and $q$ is given in terms of integers by (\ref{q}).  We note that the parameterisation of this solution has the scaling freedom
\be
\label{scaling2}
(a_0, \beta, \xi, p_0) \mapsto (t a_0, t\beta, t\xi, t p_0)
\ee
with the other constants held fixed, where $t$ is a non-zero constant. We also note that by the shift $\phi^1 \to \phi^1+ b \phi^2$ it is clear that $\omega_0$ is locally a redundant parameter. However, since by our construction the $\phi^i$ are periodic coordinates on $T^2$, we may not use these shifts to set $\omega_0$ to any value we like: we will return to this when we perform a global analysis of these metrics.

\subsubsection{Derivation of solutions}

First we observe that in general, i.e. without using the ansatz (\ref{A}) and (\ref{Gamma}), one can eliminate $|k|^2$ between (\ref{a0}) and (\ref{traceeq}) to get
\be
\label{GBdd}
(\Gamma B)'' +\frac{(2n-2)A'(\Gamma B)}{A} -2a_0 -\frac{(n-1) \Gamma |q|^2}{A^4} +6\lambda \Gamma=0 \; .
\ee
On the other hand using (\ref{A}) one finds that (\ref{abeq}) reduces to
\be
\label{GBd}
(\Gamma B)' + \frac{2(n-2) \Gamma B}{A^2} + \left( \frac{|q|^2}{2A^2} -2n +\lambda A^2 \right) \Gamma=0  \; .
\ee
Now we can eliminate $|q|^2$ from this equation using (\ref{GBdd}) to give a linear second order ODE for $\Gamma B$ whose coefficients are determined by $(A,\Gamma)$.  We find it can be written as
\be
\label{hor:Pode}
\frac{P''(x)}{x^{n-1}}= \left[ \frac{n(n-1)}{x} -\lambda(n+2) \right] \Gamma  + a_0 \; .
\ee
where
\be
P(x) \equiv \frac{x^{n-1} \Gamma B}{2} \; .
\ee
Thus, integrating (\ref{hor:Pode}) we find that $P$ is a polynomial of order $n+2$ given by (\ref{P}).
Now, (\ref{GBd}) determines $|q|^2$:
\bea \label{qsq}
B_{ij}q^iq^j &=& \frac{ 8 x \left( n -\lambda x \right) \Gamma(x) -  8 x^{-n+1} ( xP'(x)-P(x)) }{\Gamma(x)} \nonumber  \\
&=&  \frac{8}{(n+1) (\beta x+\xi)} \left[  \left( 2n \beta +\lambda\xi  -a_0 \right)x^2  + \xi(n+1) x + \frac{(n+1)p_0}{x^{n-1}} \right]
\eea
where the second equality follows from using the explicit expressions.  Also note that $|k|^2$ is determined, for example from (\ref{a0}) one gets:
\be
\label{ksq}
B_{ij}k^ik^j=  -2a_0\Gamma(x) +2\lambda \Gamma(x)^2 + \frac{2\beta P'(x)}{x^{n-1}}  - \frac{2 \beta^2 P(x)}{x^{n-1}\Gamma(x)} \; .
\ee
Note that from the definition of $P$ we have also determined
\be
B= \frac{2 P(x)}{x^{n-1} \Gamma(x)}  \; .
\ee

In order to proceed we now exploit the $GL(2,\mathbb{Z})$ automorphism group of the $T^2$ fibres of the associated principal bundles $P_{m_1, m_2}$.  As discussed in section \ref{sec:torusbundle} we may fix the $T^2$ automorphism freedom by setting $(m_1, m_2 ) \mapsto (m , 0)$. With such a choice $B_{ij}q^i q^j = B_{11} q^2$, with $q$ is given by (\ref{q}), so that (\ref{qsq}) gives
\bea
B_{11}  =  \frac{Q(x)}{q^2 x^{n-1}\Gamma(x)}
\eea
where $Q(x)$ a polynomial of order $n+1$ defined by (\ref{Q}).
Since we have already determined $B =\det B_{ij}$, it remains to deduce one more component of $B_{ij}$.

Explicitly we have
\be
B_{ij} \omega^i \otimes \omega^j = B_{11} \left[ d\phi^1+ \frac{q \bar{\sigma}}{2} + \Omega d\phi^2 \right]^2 +  \frac{B}{B_{11}} (d\phi^2)^2
\ee
where
\be
\Omega \equiv \frac{B_{12}}{B_{11}}   \; .
\ee
We can  use (\ref{00eq}) to get a first order ODE for $\Omega$. For this we use the identity (cf. ~\cite{KL1})
\be
B^{lm} B^{pq} B_{lp}' B_{mq}' = \frac{2B_{11}^2 \Omega'^2}{B}  +  \left( \frac{B_{11}'}{B_{11}} \right)^2 + \left( \frac{B_{11}'}{B_{11}} - \frac{B'}{B} \right)^2  \; .
\ee
In general (i.e. even without our ansatz) we then find
\bea
\label{omegap1}
B_{11}^2 \Omega'^2 &=& -B'' - B' \left[ \frac{(2n-2)A'}{A} +\frac{\Gamma'}{\Gamma} \right] \nonumber \\ &&+ B \left[ \frac{B_{11}'}{B_{11}} \left( \frac{B'}{B} - \frac{B_{11}'}{B_{11}} \right) - \frac{2(2n-2)A''}{A} - \frac{2\Gamma''}{\Gamma}+ \frac{\Gamma'^2}{\Gamma^2} \right]-2\lambda  \; .
\eea
We may derive another equation for $\Omega'^2$ by contracting (\ref{ijeq}) with $q_iq^j$ and using the identity
\be
q_i'B^{ij} q_j'= \frac{q^2B_{11}'^2}{B_{11}} +\frac{q^2B_{11}^3 \Omega'^2}{B}  \; .
\ee One finds (again without the ansatz)
\be
B_{11}^2 \Omega'^2 = \frac{1}{\Gamma A^{2n-2}} \left( \frac{\Gamma A^{2n-2} B B_{11}'}{B_{11}} \right)' + \frac{B_{11}(k^1+\Omega k^2)^2}{\Gamma^2} -\frac{(n-1)q^2 B_{11}}{A^4} +2\lambda  \; .
\ee
The term $k^1+\Omega k^2$ can be dealt with as follows. Notice that
\be
\label{keq}
|k|^2= B_{11}(k^1+\Omega k^2)^2 + \frac{(k^2)^2B}{B_{11}}
\ee
and therefore
\be
\label{omegap2}
B_{11}^2 \Omega'^2 = \frac{1}{\Gamma A^{2n-2}} \left( \frac{\Gamma A^{2n-2} B B_{11}'}{B_{11}} \right)' + \frac{|k|^2}{\Gamma^2}- \frac{(k^2)^2B}{\Gamma^2 B_{11}} -\frac{(n-1) q^2 B_{11}}{A^4} +2\lambda  \; .
\ee
Evaluating this for our ansatz one finds an expression for $\Omega'^2$ in general different to the one from (\ref{omegap1}).  The two agree if and only if $k^2$ is given by (\ref{k2}) and the constraint (\ref{constr1}) is satisfied.
We then find $\Omega'^2$ is of the form
\begin{equation}
 \Omega'^2 = \frac{128q^4x^{2n-2} X(x) Y(x)}{n(1+n)^2 Q(x)^4}
\end{equation} where
\begin{equation}
X(x) = n p_0 \xi (n+1) + p_0 \beta (n+1)^2 x  - v_1 \xi x^{n+1}
\end{equation}
where $v_1$ is defined by (\ref{v1}) and $Y$ is another a polynomial of order $n+1$.
Although this in principle determines $\Omega$, we will make a further assumption in order to guarantee that $\Omega$ is a rational function of $x$.

We proceed by assuming that $X,Y$ are proportional.  It can be verified that  a necessary and sufficient condition for this is
\be
\label{constraint}
-\beta v_2 p_0+n p_1 \xi v_1=0
\ee  where the constant $v_2$ is defined by (\ref{v2}).
There are a number of ways (\ref{constraint}) can be satisfied. For example, the pure Einstein case, which corresponds to $\beta=0$, $\xi=1$ and $a_0=\lambda$, automatically solves this. Instead, here we consider the ``generic" case
\be
\xi v_1 \neq 0
\ee
which allows us to solve for $p_1$:
\be
\label{p1derived}
p_1 = \frac{\beta v_2 p_0}{n\xi v_1}  \; .
\ee
With this choice we have
\begin{equation}
 \Omega'= \pm   \sqrt{\frac{2w}{n}}\; \frac{ 8q^2x^{n-1} X(x)}{(1+n) Q(x)^2}
\end{equation}
where the constants $v_3$ and $w$ are defined by (\ref{v3}) and (\ref{w}) and $w \geq 0$.
Performing the resulting integral, we find
\begin{equation}
\Omega(x)  = \omega_0 +\frac{q^2 R(x)}{Q(x)}
\end{equation}
where $\omega_0$ is an integration constant, and $R(x)$ is a polynomial which turns out to depend on whether the constant $v_4$ (\ref{v4})
vanishes or not. Here we again treat the generic case
\be
v_4 \neq 0
\ee
in which case
\be
\label{Rderived}
R(x) = \sqrt{\frac{2w}{n}}\tilde{R}(x), \qquad \text{where} \qquad \tilde{R}(x)\equiv \frac{(1+n)\beta p_0 - v_1\xi x^n}{v_4}  \; .
\ee
Notice that without loss of generality we have chosen a sign for $\Omega$ since we may always fix this using the discrete transformation $\phi^2 \to -\phi^2$.

It remains to deduce the constant $k^1$. We can use (\ref{keq}) to do this.
First note one can verify
\begin{equation}
\label{Bkid1}
B_{11} |k|^2 - (k^2)^2 B = \frac{16 v_3}{n v_1 q^2 \Gamma^2 x^{2n-2}} \left[ v_0 \tilde{R}(x)  -\frac{n \beta v_1}{8 v_4 } Q(x) \right]^2  \; .
\end{equation}
On the other hand from our various definitions we have
\be
\label{Bkid2}
B_{11}^2 (k^1+\Omega k^2)^2 = \frac{1}{\Gamma^2 x^{2n-2}} \left[ (k^1+ \omega_0 k^2) q^{-2} Q + k^2 R \right]^2  \; .
\ee
Equation (\ref{keq}) tells us that the expressions (\ref{Bkid1}) and (\ref{Bkid2}) should be equal, so we deduce
\begin{equation}
\label{k1derived}
k^1+ \omega_0 k^2 = - \frac{\epsilon\,  \text{sgn}(v_0) q\beta v_1}{2v_4}\sqrt{ \frac{v_3 n}{v_1} }  \; .
\end{equation}
which together with (\ref{k2}) determines the constant $k^1$ to be (\ref{k1}).

We now verify that the remaining equations impose no further constraints. In fact it is sufficient to check the $21$ component of (\ref{ijeq}), which in general can be written in the simple form
\be
(A^{2n-2}\Gamma B_{11}^2 \Omega')'+ \frac{k^2k_1 A^{2n-2}}{\Gamma}=0  \; .
\ee
We have determined all quantities in the above equation and it is a straightforward matter to verify that it is satisfied identically.

We emphasise that the set of solutions we have just derived are valid subject to a number of constraints on the constants: equations (\ref{constr1}), (\ref{constr2}), (\ref{constr3}) and (\ref{constr4}). We will not consider the various non-generic cases that arise when one of (\ref{constr2}, \ref{constr4}) are violated. 

\subsection{Global analysis of horizon geometry}
\label{sec:globalhor}

We now consider the problem of extending our local metrics (\ref{gammasol}) smoothly on compact manifolds. As mentioned earlier, for application to black holes, we are interested in the case $\lambda \leq 0$, so for definiteness we will restrict to this case in this section.

Positive definiteness requires $A^2 \geq 0$ and hence $x \geq 0$. Also, by definition, we require $\Gamma(x)= \beta x+ \xi>0$; hence positive-definiteness also implies $P\geq 0$ and $Q \geq 0$. Recall our solution is only valid for $\xi \neq 0$, see equation (\ref{constr2}). Suppose $\xi<0$: then, positivity of $\Gamma$ implies that $\beta \geq 0$ and therefore $\xi v_0 <0$, which contradicts (\ref{constr1}). Therefore we must have
\be
\xi >0 \; .
\ee
Possible singularities in the metric occur at $x=0, x_i, x_\mu, \infty$, where $x_i$ are any real roots of $P(x)$ and $x_\mu$ are any real roots of $Q(x)$. We will analyse the generic case with $k^2 \neq 0$ and $w>0$ -- it turns out that either $k^2=0$ or $w=0$ result in metrics which cannot be extended onto compact manifolds.

First we note that if $x=0$ is in the coordinate domain, then since the invariant $\ell^{-2}\gamma(V_1, V_1)=B_{11}=Q/(x^{n-1} \Gamma)$ must be smooth, we deduce that one of the roots of $Q$ must be at $x=0$. Hence using the explicit form for $Q$ given by (\ref{Q}) we deduce $p_0=0$. It then follows that $p_1=0$ from (\ref{p1}).  We treat this case in Appendix \ref{app:MP} and show that smoothness implies this horizon has $S^{2n+1}$ topology. Furthermore, we show it is equivalent to a class of Myers-Perry near-horizon geometries (at least for $\Lambda=0$).  Hence we will not discuss this further.

Hence we may now assume that $x>0$.  By examining the invariant $\ell^{-2} \gamma(V_2, V_2)=B_{22}$ it is easy to see that smoothness requires that if a root of $Q(x)$ is in our domain, then it must also be a root of $2x^{n-1}\Gamma P+R^2$. Because each term is non negative we deduce that any such root of $Q$ must be a root of both $P$ and $R$.  To analyse this case, we note the following general identity, which may be directly verified:
\be
\frac{ \beta (n+1)Q(x)}{8} = v_4\left[  x^n \Gamma(x) + \sqrt{\frac{n}{2w}} R(x) \right]  \; .
\ee
From this is follows that at any simultaneous root of $Q$ and $R$ we must have $\Gamma=0$, which contradicts our requirement that $\Gamma>0$. Hence it follows that any root of $Q$ cannot be in our domain.

Hence we may assume that $Q>0$ in our domain. Compactness then requires that $0<x_1\leq x \leq x_2$ where $x_1<x_2$ are two adjacent simple roots of $P$. Generically our solutions will have $\Omega' \neq 0$ for all $x \in [x_1,x_2]$, or at least $\Omega(x_1)\neq \Omega(x_2)$. Our metric is then exactly of the form considered in section \ref{sec:global}, where conditions were determined in order to extend the metric smoothly onto a total space $S$ which is a Lens space bundle over $K$. The resulting smoothness conditions for the present case can be written as
\be \label{hor:regconds}
\frac{m_i}{j} = X_i \equiv \left(\frac{p}{2n}\right) \left( \frac{ G(x_i)}{ \frac{R(x_2)}{Q(x_2)}-\frac{R(x_1)}{Q(x_1)}} \right)  \qquad \text{for} \quad i=1,2 \;,
\ee
where $m_i$ are the non-zero integers specifying the associated $T^2$-bundle $P_{m_1,m_2}$ and $j$ is a non-zero integer specifying the Lens space fibre $L(j,k)$.  If solutions to (\ref{hor:regconds}) exist, i.e. integers $m_i,j$ exist such that (\ref{hor:regconds}) can be satisfied, then our metrics extend to smooth metrics on the Lens space bundles $S$.

\par We now discuss existence of solutions to these regularity conditions.  Since $\omega_0$ can be fixed, our local metrics (\ref{gammasol}) are parametrised by the five continuous parameters $(\ell, a_0, \beta, \xi, p_0)$. The remaining scaling freedom (\ref{scaling2}) can be used to fix one further parameter, which for concreteness we take to be $a_0$. We are therefore left with four parameters $(\ell, \beta, \xi, p_0)$. These are subject to the inequalities satisfied by our local metrics discussed earlier, together with further inequalities required for the existence of two positive adjacent simple roots $x_i$ of $P$ such that $\Gamma>0,Q>0, P \geq 0, \Omega' \neq 0$ for all $x \in [x_1,x_2]$.  Existence of solutions then reduces to showing that the two functions $X_i$ on the RHS of (\ref{hor:regconds}) take rational values. It is evident that this is the case, as these are continuous function of four parameters. Hence, there exists a countably infinite set of rational solutions to the two conditions (\ref{hor:regconds}), each of which generically depends on two combinations of the continuous parameters (the other two combinations are fixed in terms of the integers by (\ref{hor:regconds})). This establishes the existence of an infinite class of smooth horizon manifolds and hence our theorem \ref{theorem1}.  In particular, note that by varying $j$ it is clear there exists a countably infinite set of non-zero integers $(m_1,m_2)$ which solve the regularity conditions.  

It would be interesting to determine more explicitly the allowed ranges of these continuous functions $X_i$, in order to determine the precise set of integers $(m_1,m_2,j)$ that are allowed; unfortunately this appears to be rather complicated due to the number of parameters involved and determining their explicit domain.

\subsection{Sasakian horizons}
\label{sec:sasakihorizon}

The class of explicit solutions found above includes a case with $\Gamma$ equal to a constant. Note that for us this is equivalent to $h$ being a Killing vector field. Recently a set of solutions with this property were found~\cite{KL4}.  In fact, as we now show, these arise as the special case of our solution given by $\beta=0$ and $p_1=0$.

Since $\Gamma=\xi$  and $\Gamma$ is only defined up to a scaling (\ref{scaling2}), without loss of generality we also set $\xi=1$ so $\Gamma=1$. Then, as a vector field, $h= k^i \frac{\partial}{\partial \phi^i}$. In this case the polynomials $P,Q,R$ simplify somewhat and are given by
\bea \label{Sfunc}
P(x) &=& p_0+ x^n +\frac{[a_0- \lambda (n+2)]}{n(n+1)} x^{n+1}   \\
 Q(x) &=& \frac{8}{(n+1)}[ (\lambda-a_0) x^{n+1}+(n+1)(x^n+p_0)]   \\
 R(x) &=&- \sqrt{ \frac{2(2\lambda-a_0)}{n}}  x^n
\eea
and the various constraints on the parameters simplify to just
\be
a_0 < \lambda \qquad \qquad  a_0 < 2\lambda  \; .
\ee
The constants $k^i$ are given by
\be
k^2 = \pm \frac{2\sqrt{2(\lambda-a_0)}}{q}  \qquad  \qquad k^1=- \omega_0 k^2  \; .
\ee
Notice from (\ref{ksq}) we see that
\be
|k|^2=2(\lambda-a_0)
\ee
which allows us to eliminate the constant $a_0$ in favour of $|k|$, in particular
\be
P(x)= p_0+ x^n - \frac{\hat{\lambda} x^{n+1}}{n+1}
\ee
where
\be
\hat{\lambda} =  \frac{|k|^2 + 2\lambda(n+1)}{2n} \; .
\ee
By recompleting the square we may simplify the matrix $B_{ij}$:
\be
B_{ij} \omega^i \omega^j  =  \left[ d\hat{\phi}^2 -2 x \sqrt{ \frac{\frac{|k|^2}{2}+\lambda}{2n}} \left(d\hat{\phi}^1+ \bar{\sigma} \right) \right]^2 + \frac{2P(x)}{x^{n-1}} \left(d\hat{\phi}^1 + \bar{\sigma} \right)^2
\ee
where for convenience we have defined new coordinates
\be
\hat{\phi}^1= \frac{2}{q}( \phi^1+\omega_0 \phi^2) \qquad \qquad \hat{\phi}^2=\frac{q}{2} \phi^2
\ee
and made use of the identity
\be
\label{Sid}
2P(x)+ \frac{R(x)^2}{x^{n-1}} = \frac{Q(x)}{4}
\ee
which is valid for $\beta=p_1=0$. In these coordinates we simply have
\be
h= \pm |k| \frac{\partial}{\partial \hat{\phi}^2}  \; .
\ee
It is now clear this is precisely the solution recently presented in~\cite{KL4}. To get from the solution in~\cite{KL4} to the one presented here one must perform a dilation $\gamma_{AB} \to \ell^2 \gamma_{AB}$ with $h_A \to h_A$, which leaves the horizon equation invariant apart from the replacement $\Lambda \to \Lambda  \ell^2=\lambda$.  A global analysis of this solution was performed in~\cite{KL4} using the same method as in~\cite{GMSW}.

We may recover these results as a special case of the general case analysed in section \ref{sec:global}.  Using the identity (\ref{Sid}) allows one to show the quantities $X_i$ defined in (\ref{hor:regconds}) simplify to
\be
X_1 =  \left(\frac{p}{n}\right) \frac{x_2 ( n- \hat{\lambda} x_1)}{x_2-x_1} \; ,  \qquad \qquad
X_2=  -\left(\frac{p}{n}\right) \frac{x_1 ( n- \hat{\lambda} x_2)}{x_2-x_1}
\ee
and therefore $X_1+X_2=p$. Hence
\be
\label{sumsi}
m_1+m_2=jp
\ee
and
\be
\frac{m_1}{m_2} =\frac{x_2(n-\hat{\lambda} x_1)}{x_1(\hat{\lambda} x_2-n)}  \; .
\ee
These regularity conditions are in the same form as the ones for the Sasaki-Einstein manifolds given in~\cite{MS}.  In fact using the ranges of the $x_i$  given in~\cite{KL4} we may deduce that the function $X_1$ is a monotonic function with range on the interval $p/2<X_1<p$. Hence there exists an infinite number of solutions $m_1/j$ such that $p/2< m_1/j<p$, just as in the Sasaki-Einstein case~\cite{MS}.

Consider the five dimensional case $n=2$, which must have $p=2$.  The condition (\ref{sumsi}) implies we can write $m_1=j+l$ and $m_2=j-l$ for some $l \in \mathbb{Z}$. Hence $S$ is the contact quotient $S^7 // (j+l, j-l, -j, -j )$ which shows that it must be diffeomorphic to $S^3\times S^2$~\cite{MStoric}.

We note that we can recover the analysis given in~\cite{KL4} if we set $t_1=-t_2=-1$. This case is of the type discussed earlier in section \ref{sec:U1bundle}. Using the identity (\ref{Sid}), allows one to show
\be
\epsilon_i G(x_i)+ \frac{2n R(x_i)}{Q(x_i)} = \frac{a_0 -\lambda(n+2)}{2 \sqrt{2n(2\lambda-a_0)}}
\ee
where $\epsilon_1=1$ and $\epsilon_2=-1$, at each root $x_i$ of $P(x)$. Then, (\ref{app:ti}) implies
\be
m=p  \label{sasaki:m}
\ee
and
\be
\label{sasaki:omega0}
\omega_0 = - \sqrt{\frac{n}{2(2\lambda-a_0)}} \;  [a_0 -\lambda(n+2)]  \; .
\ee
Also note that the fibre which has Lens space topology $L(j,k)$ simplifies. We have $j=s_1+s_2$ and $k= \hat{t}_1+ s_2 \hat{s}_1$ where $-\hat{t}_1 + s_1 \hat{s}_1=1$, so $k=-1+ \hat{s}_1 j \equiv -1 \; \text{mod} \, j$.


\subsection{Topology of horizons}

We have constructed an infinite class of horizon metrics with $H \cong S$, i.e. cross-sections of the horizon are Lens bundles over a Fano K\"ahler-Einstein manifold $K$. The topology of these manifolds is discussed in section \ref{sec:topology}, \ref{sec:global} and \ref{sec:special}. In particular, if $\text{gcd}(m_i,j)=1$ for both $i=1,2$, our five dimensional horizon manifolds are all diffeomorphic to $S^3\times S^2$ or $S^3 \tilde{\times} S^2$ depending on whether $m_1+m_2$ is even or odd respectively.

For application to black hole solutions we require that the topology of $H$ must be positive Yamabe type for $\Lambda \geq 0$. In fact this is guaranteed for any non-trivial solution to (\ref{Heq}), as pointed out in~\cite{Lucietti}.

For asymptotically flat or globally AdS black holes one also requires that $H$ is oriented-cobordant to $S^{2n+1}$. It is clear that this is the case for our five dimensional horizon manifolds ($n=2$) . Furthermore, any compact oriented $7$-manifold is cobordant to $S^7$. By an identical argument as in~\cite{KL3}, it is easy to see that any $S^3$-bundle is the boundary of some compact manifold (namely the associated ball bundle), and therefore these all oriented-cobordant to $S^{2n+1}$.  For the general case of Lens space bundles over $K$, we may argue as follows. Construct the associated bundles to $S$ formed by replacing the $L(j,k)$-fibres with a toric resolution of the cone over $L(j,k)$. The manifold $S$ is then the boundary of the compact manifold defined by taking the interior of the link of the cone. Hence it must be oriented-cobordant to a sphere.

Hence all our examples satisfy the known constraints on topology for asymptotically flat and globally AdS black holes.

\subsection{Physical quantities}

The area of our horizon geometries is given by
\be
A(H) = \int_H \sqrt{\gamma}= \frac{2^{n+1} \pi^2 c \, \ell^{2n+1} (x_2^n-x_1^n) V_K}{n}
\ee
where $V_K=\int_K \sqrt{\bar{g}}$ is the volume of $K$.
The Komar angular momentum with respect to a Killing field $\xi$ can be computed from the near-horizon geometry using~\cite{FKLR}
\be
J[\xi]  = \frac{1}{16 \pi} \int _H \sqrt{\gamma} \; h \cdot \xi \; .
\ee
For our case we find
\bea
J[\partial_{\phi_1}] &=& \frac{2^{n-1}  \pi c\, \ell^{2n+1} V_K}{4q^2} \int_{x_1}^{x_2}dx\; \frac{Q}{\Gamma} \left( k^1+\omega_0 k^2 +\frac{q^2 R k^2}{Q} \right) \nonumber  \\
&=& -\frac{2^{n-1} \epsilon\, \text{sgn}(v_0) \pi c\, \ell^{2n+1} V_K}{q(1+n)}\sqrt{\frac{v_3}{v_1 n}} \left[ \frac{(1+n)p_0 + v_1 x^{1+n}}{\Gamma(x)} \right]^{x_2}_{x_1}
\eea
and
\bea
J[\partial_{\phi_2}-\omega_0 \partial_{\phi_1}] &=& \frac{2^{n-1} \pi c \, \ell^{2n+1}V_K}{4} \int_{x_1}^{x_2} dx \; \left[ \frac{R }{ \Gamma^2}\left(k^1 + \omega_0 k^2+ \frac{q^2R k^2}{Q} \right)+ \frac{2q^2 x^{n-1} P k^2}{\Gamma Q}\right] \nonumber \\
&=&  \frac{2^{n-1} \epsilon q \pi c \, \ell^{2n+1}V_K}{4n v_1 v_4}  \sqrt{\frac{v_0}{2 \xi}}\left[\frac{\xi v_1 x^n( v_4 - \lambda \beta x) - \beta p_0(1+n) v_3}{\Gamma(x)}\right]^{x_2}_{x_1}  \; .
\eea
Therefore, generically both angular momenta in the $T^2$-fibre are non-vanishing. Using (\ref{sasaki:m}) and (\ref{sasaki:omega0}) it can be verified that the Sasakian special case discussed in section \ref{sec:sasakihorizon},  reduces to $J[\partial_{\phi_2}] =0$, in agreement with~\cite{KL4}. Hence our new near-horizon geometries may be thought of as ``doubly-spinning" versions of the Sasakian ones.  \\

\noindent {\bf Acknowledgements}
HK is supported by an NSERC Discovery Grant. JL is supported by an EPSRC Career Acceleration Fellowship. We would especially like to thank James Sparks for explaining to us a number of aspects of the topology of the manifolds considered in this paper. HK would also like to thank Tom Baird and Eduardo Martinez-Pedroza for useful discussions.


\appendix

\section{Curvature computations}
\label{app:curvature}
To perform calculations in this paper it is convenient to use the following non-coordinate (non-orthonormal) basis of 1-forms $e^A$, where $A=0, i, a$ and $i=1,2$ and $a=1, \dots 2n-2$,
\be
e^0 = \frac{\ell dx}{\sqrt{B(x)}}, \qquad   \qquad e^i = \ell \left( d\phi^i+ \frac{q^i \bar{\sigma}}{2} \right) , \qquad  \qquad e^a =\ell \bar{e}^a  \; ,
\ee
where $\bar{e}^a$ are vielbeins for $\bar{g}$,
so that the metric
\be
\gamma= e^0 \otimes e^0 + B_{ij}(x) e^i \otimes e^j +A^2(x) \delta_{ab} e^a \otimes e^b  \; .
\ee
The dual basis vector fields $e_A$ are given by
\be
e_0 = \frac{\sqrt{B(x)}}{\ell} \frac{\partial}{\partial x}, \qquad \qquad e_i =\frac{1}{\ell} \frac{\partial}{\partial \phi^i}, \qquad \qquad e_a =\frac{1}{\ell} \left( \bar{e}_a - \frac{q^i \bar{\sigma}_a}{2} \frac{\partial}{\partial \phi^i} \right)
\ee
where $\bar{e}_a$ are the dual vectors to $\bar{e}^a$.  For convenience of notation we define $B^{ij}$ as the components of the inverse matrix with components $B_{ij}$, $q_i =B_{ij} q^j$ and $|q|^2 = B_{ij}q^iq^j$.

As is standard we define the components of the Levi-Civita connection of the metric $\gamma$ by $\nabla_A e_B= \Gamma^C_{BA} e_C$. In a general basis these are given by
\be
 \Gamma^D_{BC} = \frac{1}{2}\gamma^{AD} \left[ e_B(\gamma_{AC})+ e_C(\gamma_{BA}) -e_A(\gamma_{CB}) + c_{ACB} +c_{BAC}-c_{CBA} \right]
\ee
where $[e_A, e_B] =c^C_{~~AB} e_D$ and $c_{ABC}= \gamma_{AD}c^D_{~~BC}$. For the case at hand we find that the only non-vanishing structure constants are
\be
c^a_{~~bc} = \frac{\bar{c}^a_{~~bc}}{\ell} \qquad \qquad c^i_{~~ab} = -\frac{q^i \bar{J}_{ab}}{\ell}
\ee
where $\bar{c}^a_{~~bc}$ are the structure constants of the frame $\bar{e}_a$. We find that the non-vanishing components of $\Gamma^A_{~~BC}$ in this basis are:
\bea
&&\Gamma^0_{~ij} = -\frac{\sqrt{B} B_{ij}'}{2\ell} \qquad \Gamma^0_{~~ab}= -\frac{\sqrt{B} (A^2)'}{2\ell}\delta_{ab} \qquad \Gamma^{i}_{~~0j} =\frac{\sqrt{B} B^{ik} B_{kj}'}{2\ell} \\ &&\Gamma^i_{~ab} = \frac{q^i \bar{J}_{ab}}{2\ell}  \qquad \Gamma^a_{~~0b} = \frac{\sqrt{B} A' \delta^a_{~b}}{\ell A} \qquad \Gamma^a_{~ib} = -\frac{q_i}{2\ell A^2} \bar{J}^a_{~b}\qquad  \Gamma^a_{~~bc} = \frac{1}{\ell} \bar{\Gamma}^a_{~~bc}  \; .
\eea

In a general basis the Ricci tensor can be computed directly from
\be
R_{BD} = e_A(\Gamma^A_{~BD})- e_D(\Gamma^A_{~BA}) +\Gamma^C_{~BD}\Gamma^A_{~CA}- \Gamma^C_{~BA}\Gamma^A_{~CD} - c^C_{~AD}\Gamma^A_{~BC}  \; .
\ee
From this one can check that the Ricci curvature is:
\bea\label{Ricci}
\ell^2 R_{00}&=& -\frac{\sqrt{B}(A' \sqrt{B})'(2n-2)}{A} - \sqrt{B} (\sqrt{B})'' - \frac{B B^{lm}B^{pq}}{4} B_{lp}' B_{mq}'
\\
\ell^2 R_{ij} &=& (n-1) \left[ \frac{q_iq_j}{2A^4} -\frac{B A' B_{ij}'}{A} \right] - \frac{B' B_{ij}'}{2} + \frac{B}{2} \left[ -B_{ij}'' + B^{lm} B_{il}' B_{jm}' \right] \\
\ell^2 R_{ab}&=& \bar{R}_{ab} - \delta_{ab}\left[\frac{|q|^2}{2A^2} + A'^2 B (2n-3)+ A \left( BA' \right)' \right]   \; .
\eea

The  horizon equation (\ref{Heq}) can be written as $R_{AB}=S_{AB}$ where the``source" term is
\be
S_{AB}= \frac{1}{2} h_A h_B -\nabla_{(A} h_{B)}+\Lambda \gamma_{AB}  \; .
\ee
In our basis
\be
h= -\frac{\sqrt{B} \Gamma'}{\ell \Gamma} e^0 + \frac{k_i}{\ell \Gamma} e^i  \; ,
\ee
and recalling $\nabla_A h_B= e_A(h_B)-\Gamma^D_{~BA}h_D$, we find the following non-vanishing components:
\bea
&&\ell^2 S_{00} = \frac{\sqrt{B}(\sqrt{B} \Gamma')'}{\Gamma} -\frac{B\Gamma'^2}{2\Gamma^2} +\lambda ,  \qquad  \qquad \ell^2S_{ab}  = \left(\frac{B \Gamma' (A^2)'}{2\Gamma}+ A^2 \lambda \right) \delta_{ab}, \\
&&\ell^2 S_{ij} = \frac{k_ik_j}{2\Gamma^2} + \frac{B \Gamma' B_{ij}'}{2\Gamma} +\lambda B_{ij} , \qquad  \qquad \ell^2 S_{0i} = - \frac{\sqrt{B} B_{ij}(k^j)'}{2 \Gamma},
\eea
where we have defined $\lambda= \Lambda \ell^2$ and $k^i= B^{ij}k_j$.

Using the above we see that the $00$ and $ab$ components of the horizon equation (\ref{Heq}) give the ODEs (\ref{00eq}) and (\ref{abeq}) respectively (for the latter we also used $\text{Ric}(\bar{g}) = 2n \bar{g}$).  The $ij$ component gives
\be
B( B_{ij}'' - B^{lm}B_{il}'B_{jm}')+ B' B_{ij}' +2(n-1) \left[ \frac{B A' B_{ij}'}{A} - \frac{q_iq_j}{2A^4}\right] +\frac{k_ik_j}{\Gamma^2} + \frac{B \Gamma' B_{ij}'}{\Gamma} +2\lambda B_{ij}=0 \; .
\ee
This may be simplified a little by multiplying with the inverse matrix $B^{ki}$ and defining
\be
C^k_{~j} = B^{ki} B_{ij}'
\ee
which results in the ODE (\ref{ijeq}).

Finally it is also worth recording the near-horizon function (\ref{Feq}) which for us is
\be
\ell^2 F= \frac{(B\Gamma')'}{2\Gamma} +\frac{|k|^2}{2\Gamma^2} +\frac{(n-1) B A' \Gamma'}{A \Gamma} +\lambda  \; .
\ee

\section{Spherical topology horizons}
\label{app:MP}

In this Appendix we show that setting $p_0=p_1=0$ in our solution (\ref{gammasol}), (\ref{hsol}), and demanding that the horizon metric extends smoothly onto a compact manifold, results in a spherical horizon topology $H= S^{2n+1}$.  We also show that the resulting near-horizon data corresponds to that of the extremal Myers-Perry black hole in $2n+3$ dimensions with all but one angular momenta equal and non-zero. Therefore, within our class of solutions, we do not find any spherical topology horizons more general than the Myers-Perry horizons.

\subsection{Special case of near-horizon geometry}
Set $p_0=p_1=0$ in (\ref{gammasol}). The near-horizon data simplifies to
\be
\ell^{-2} \gamma = \frac{\Gamma dx^2}{2x \hat{P}} + \frac{x \hat{Q}}{q^2 \Gamma}\left[d\phi^1+\frac{q  \bar{\sigma} }{2}+\left(\omega_0 +  \frac{q^2 \hat{r}}{\hat{Q}}\right) d\phi^2\right]^2 + \frac{2q^2 \hat{P}(d\phi^2)^2}{\hat{Q}} + 2x \bar{g}
\ee
where
\bea
&&\hat{P}(x) \equiv \frac{P(x)}{x^n}  = -\frac{\beta \lambda x^{2} }{(n+1)}  + \frac{\left[ n(n-1) \beta+ a_0 -\lambda \xi (n+2) \right]x}{n(n+1)}   +\xi  \\
&&\hat{Q}(x) \equiv \frac{Q(x)}{x^n}  =  \frac{8}{ (n+1)} \left[  \left( 2n \beta +\lambda\xi  -a_0 \right)x  + \xi(n+1)\right]  \\
&&  \hat{r} \equiv \frac{R(x)}{x^n}  =- \sqrt{\frac{2w}{n}}\frac{v_1\xi }{(2n\beta - a_0 + \xi \lambda)}  \; .
\eea

There are potential singularities at $x=0$ and any root of $\hat{P}$. First consider this metric near $x=0$ by setting $x= \rho^2$ and expand for small $\rho$. One gets
\be
\ell^{-2}\gamma= 2 d\rho^2 + 2 \rho^2 \left[ \frac{4}{q^2}\left( d\phi^1+\frac{q \bar{\sigma}}{2}\right)^2+ \bar{g} \right] +O(\rho^2)\left( d\phi^1+\frac{q \bar{\sigma}}{2}\right) d\phi^2+ \frac{q^2}{4} (d\phi^2)^2+ \dots
\ee
where $\dots$ signify higher order terms in $\rho$. Hence, since $\Delta\phi^1=2\pi$, in order to avoid a singularity at $x=0$ we must have $q=2$ and furthermore $\bar{g}$ must be the Fubini-Study metric on $\mathbb{CP}^{n-1}$, and so $p=n$ and $m=1$. The space near $x=0$ approaches $\mathbb{R}^{2n}\times S^1$. We note that since $\Gamma>0$ we must have $\xi>0$.

Now consider the metric near a root $x_0>0$ of $\hat{P}$.  The Killing field $\partial / \partial \phi^2- \Omega(x_0) \partial /\partial \phi^1$ vanishes at $x=x_0$ on codimension-2 submanifolds. Smoothness requires this vector to have closed orbits and since by construction $\Delta \phi^1=2\pi $ and $\Delta \phi^2=2\pi c$, we must have $\Omega(x_0) c \in \mathbb{Z}$. Hence we may use the freedom $\phi^1\to \phi^1 +b \phi^2$ to set $\omega_0 = -q^2 \hat{r}/ \hat{Q}(x_0)$. Then, the Killing vector field $\partial/\partial \phi^2$
vanishes at $x=x_0$. Requiring that the associated conical singularity is absent is equivalent to
\be
c^2 = \frac{\Gamma(x_0) Q(x_0)}{4 x_0 P'(x_0)^2}  \; .
\ee
Near $x=x_0$ our space is then smooth with topology $\mathbb{R}^2 \times S^{2n-1}$.

Hence with these choices on the parameters, the above horizon metric extends smoothly to $0 \leq x \leq x_0$, giving a metric on $S^{2n+1}$. This in fact is the horizon geometry of the extremal Myers-Perry black holes with all but one angular momenta equal, which for convenience is given in the next section.

 We will verify this in the simpler case of $\lambda=0$. First note that $\hat{P}$ becomes a linear function and, since $\xi>0$, existence of the root $x_0>0$ requires  $a_0 < -n(n-1)\beta$. Now use the scaling (\ref{scaling2}) to arrange $x_0=1$. This requires
 \be
 a_0 = - n[ (n-1)\beta + (n+1) \xi]  \; ,
 \ee
 which implies
 \be
 c = \frac{\sqrt{2}}{\xi }\sqrt{ (\beta+\xi)[ n\beta+ (n+1)\xi]}  \; .
 \ee
Define $\tilde{\phi}^2 = -c^{-1} \phi^2$ so that $\Delta \tilde{\phi}^2= 2\pi$ (the minus sign is for later convenience).
The inner products of the Killing fields are
\bea
\ell^{-2} \left| \frac{\partial }{\partial \phi^1} \right|^2 &=& 2x+ \frac{2x^2 [ (n-1)\beta+ n \xi]}{\Gamma}  \label{MP11}\\
\ell^{-2} \frac{\partial}{\partial \phi^1} \cdot  \frac{\partial}{\partial \tilde{\phi}^2}  &=&\frac{2(\beta+\xi)}{\Gamma} \sqrt{\frac{n[(n-1)\beta+n \xi]}{\xi}} \; x(1-x) \\
\ell^{-2} \left| \frac{\partial }{\partial \tilde{\phi}^2} \right|^2 &=& \frac{2(1-x)(\beta+\xi)}{\xi} + \frac{2n(1-x)^2 (\beta+\xi)^2}{\xi \Gamma}  \label{MP22}
\eea
and the $xx$ part of the metric is just
\be
\ell^{-2}\gamma_{xx}=\frac{\Gamma dx^2}{2\xi x(1-x)} \; .
\ee
Notice that the metric now depends only on $\beta/\xi$  and the overall scale $\ell^2$. The rest of the near-horizon data also simplifies and is given by
\be
k^2 = \pm \sqrt{2n\xi [n\beta+(n+1)\xi]},  \qquad \qquad k^1 = \mp \sqrt{ [(n-1)\beta+ n\xi] (\beta+\xi)}
\ee
where the signs are correlated (i.e. the $k^i$ have opposite signs). Note that $k^2$ here is with respect to $\phi^2$; converting to $\tilde{\phi}^2$ then gives
\be
\tilde{k}^2 = \mp \xi \sqrt{ \frac{n \xi}{\beta+\xi}}  \; .
\ee

We are now in a position to compare to the Myers-Perry horizon derived in the next section. Let $(r_+, a_1, a_2)$ be three positive constants satisfying (\ref{r+}). Let
\be
\ell^2 = \frac{r_+^2+a_1^2}{2}, \qquad \qquad  \frac{\beta}{\xi} = \frac{a_2^2-a_1^2}{r_+^2+a_1^2} \; .
\ee
It is then immediate that the $\bar{g}$ and $xx$ part of our above metric agrees with that of MP.  Furthermore the inner products of the Killing fields agree with those of MP as a consequence of (\ref{r+}).  Similarly, one can check that $k^1/k^2$ also agrees with that of MP as a consequence of $(\ref{r+})$.  Finally, by defining $\xi$ suitably one can arrange that one of the $k^i$ agrees -- this then shows the two near-horizon geometries are isometric. Using (\ref{r+}) one can write it in the symmetric form
\be
\xi = \frac{2r_+^2}{(r_+^2+a_1^2)(r_+^2 +a_2^2)}\;.
\ee

\subsection{Myers-Perry horizons}
Consider the $D=2n+3$ extremal Myers-Perry black hole with $a_2 \neq a_1$ and $a_i=a_1$ for $i=3, \dots n+1$. Without loss of generality we assume $a_1>0$ and $a_2>0$. Its near-horizon geometry was calculated in~\cite{FKLR} and is given by\footnote{As compared to~\cite{FKLR}, we have swapped $a_1$ and $a_2$ in order to be consistent with the conventions  in this paper. Also note there is a term missing from the horizon metric in that paper.}:
\bea
\gamma &=& \rho^2(\theta) d\theta^2 + (r_+^2+a_2^2)\cos^2\theta d\psi^2 + (r_+^2+a_1^2)\sin^2\theta [ (d\phi+\bar{\sigma})^2 +\bar{g}_{n-1} ] \nonumber \\
&+&\frac{(r_+^2+a_1^2)(r_+^2+a_2^2)}{r_+^2 \rho^2(\theta)} \left[ a_2 \cos^2\theta d\psi +a_1 \sin^2\theta (d\phi+\bar{\sigma}) \right]^2   \\
h &=& \frac{1}{\Gamma} \left( k_{\phi^i} d\phi^i - d\Gamma \right) \\
 \Gamma &=& \frac{r_+^2 \rho(\theta)^2}{(r_+^2 +a_1^2)(r_+^2+a_2^2)}  \qquad \qquad k^{\phi^i} = \frac{2r_+a_i}{(r_+^2+a_i^2)^2}
\eea
where
\be
\rho^2(\theta) =r_+^2 +a_1^2 \cos^2\theta+a_2^2 \sin^2\theta
\ee
and $0 \leq \theta \leq \pi/2$, $\phi=\phi^1$, $\psi=\phi^2$ are both $2\pi$ periodic and $\bar{g}_{n-1}$ is the Fubini-Study metric on $\mathbb{CP}^{n-1}$ with K\"ahler form $\bar{J}=\frac{1}{2}d\bar{\sigma}$. The metric is cohomogeneity-1 and parameterized by $(a_1,a_2)$, where $r_+$ is given by the largest root of
\be
\label{r+}
nr_+^4 +(n-1) a_2^2 r_+^2 -a_1^2a_2^2=0
\ee
We will now rewrite this near-horizon geometry in the coordinates used in this paper.

Define a new coordinate $0 \leq x \leq 1$ by
\be
x= \sin^2\theta  \; .
\ee
It follows that
\bea
\gamma &=&  \frac{(r_+^2 + a_1^2 + (a_2^2-a_1^2)x) dx^2}{4 x(1-x)}  + (r_+^2+a_2^2) (1-x) d\psi^2 + (r_+^2+a_1^2) x \left[ (d\phi+\bar{\sigma})^2+ \bar{g}_{n-1}\right] \nonumber \\ &&+\frac{1}{\Gamma(x)}\left[ a_2 (1-x) d\psi +a_1 x (d\phi+\bar{\sigma}) \right]^2 \\
&& \Gamma(x) = \frac{r_+^2 [r_+^2 + a_1^2 + (a_2^2-a_1^2)x ]}{(r_+^2+a_1^2)(r_+^2+a_2^2) }   \; .
\eea
This horizon data is in the general form considered in this paper (\ref{gamma}) and (\ref{h}). To see this we must identify the matrix $B_{ij}$ which in this case is given by
\be
\ell^2 B_{ij} \omega^i \omega^j = (r_+^2+a_1^2) x \, (\omega^1)^2+ (r_+^2+a_2^2) (1-x)\, (\omega^2)^2 + \frac{1}{\Gamma(x)}\left[ a_1 x \, \omega^1+ a_2 (1-x)\, \omega^2 \right]^2
\ee
from which it follows that
\be
B= \det B_{ij} = \frac{(r_+^2+a_1^2)(r_+^2+a_2^2) x (1-x)}{\ell^4\Gamma(x)}
\ee
and hence we see that $\gamma_{xx} = \ell^2/ B$ where
\be
\ell^6 = \frac{(r_+^2+ a_1^2)^2 (r_+^2+a_2^2)^2}{4 r_+^2 }
\ee
is a  constant. We also see that
\be
\ell^2 A(x)^2 = (r_+^2+a_1^2) x
\ee
and hence this class of solutions falls inside the local ansatz used in this paper.


\end{document}